\documentclass{ieeeaccess}

\pdfoutput=1

\usepackage{color}
\usepackage{pifont}
\usepackage[usenames,dvipsnames]{xcolor}

\usepackage{cite}
\usepackage{amsmath,amsfonts}
\usepackage{amssymb}

\usepackage{mathtools}
\usepackage{multirow}
\usepackage{comment}

\usepackage{algorithm}

\usepackage{algpseudocode}
\usepackage{listings}

\usepackage{mdframed}

\usepackage{graphicx}
\usepackage{textcomp}

\def\BibTeX{{\rm B\kern-.05em{\sc i\kern-.025em b}\kern-.08em
    T\kern-.1667em\lower.7ex\hbox{E}\kern-.125emX}}
    
 \usepackage{pgfplots}
\usepackage{pgfplotstable}
\pgfplotsset{compat=1.7}
\usepackage{tikz}
\usetikzlibrary{patterns}

\NewSpotColorSpace{PANTONE}
\AddSpotColor{PANTONE} {PANTONE3015C} {PANTONE\SpotSpace 3015\SpotSpace C} {1 0.3 0 0.2}
\SetPageColorSpace{PANTONE}

\begin{document}
\doi{10.1109/ACCESS.2020.3009748}
\date{\phantom{x}}
\title{HARMer: Cyber-attacks Automation and Evaluation}
\author{\uppercase{Simon Yusuf Enoch}\authorrefmark{1}, \authorrefmark{3},
\uppercase{Zhibin Huang\authorrefmark{1}, \uppercase{Chun Yong Moon}\authorrefmark{1}, \uppercase{Donghwan Lee}\authorrefmark{2}, \uppercase{Myung Kil Ahn}\authorrefmark{2,4}, and Dong Seong Kim}.\authorrefmark{1}}
\address[1]{School of Information Technology and Electrical Engineering, The University of Queensland, Brisbane, QLD 4072, Australia (e-mail: sey19@uclive.ac.nz, zhibin.huang@uq.net.au, c.moon@uq.edu.au and dan.kim@uq.edu.au)}
\address[2]{The 2nd R\&D Institute - 3rd Directorate, Agency for Defense Development, Ogeum-ro 460, Songpa-gu, Seoul, Rep. of Korea (e-mail: dlee@add.re.kr,  happyahn@add.re.kr)}
\address[3]{Department of Computer Science, Federal University Kashere, Gombe, Nigeria}
\address[4]{School of EEE, Chung-Ang University, Seoul 06974, Korea, (e-mail:lovedew@cau.ac.kr)}
\tfootnote{This work was supported by a grant (U19059EF) from the Agency of Defence and Development, Republic of South Korea.} 

\markboth
{Enoch \headeretal: HARMer: Cyber-attacks Automation and Evaluation}
{Enoch \headeretal: HARMer: Cyber-attacks Automation and Evaluation}

\corresp{Corresponding author: Simon Yusuf Enoch (e-mail: e.yusuf@uq.edu.au).}

\begin{abstract}

With the increasing growth of cyber-attack incidences, it is important to develop innovative and effective techniques to assess and defend networked systems against cyber attacks. One of the well-known techniques for this is performing penetration testing which is carried by a group of security professionals (i.e, red team). Penetration testing is also known to be effective to find existing and new vulnerabilities, however, the quality of security assessment can be depending on the quality of the red team members and their time and devotion to the penetration testing.

In this paper, we propose a novel automation framework for cyber-attacks generation named `HARMer' to address the challenges with respect to manual attack execution by the red team. Our novel proposed framework, design, and implementation is based on a scalable graphical security model called Hierarchical Attack Representation Model (HARM).  (1) We propose the requirements and the key phases for the automation framework. (2) We propose security metrics-based attack planning strategies along with their algorithms. (3) We conduct experiments in a real enterprise network and Amazon Web Services. 
The results show how the different phases of the framework interact to model the attackers' operations. This framework will allow security administrators to automatically assess the impact of various threats and attacks in an automated manner.

\end{abstract}


\titlepgskip=-15pt

\maketitle

\section{Introduction}
\label{sec:introduction}

Despite the billions of dollars spent on the prevention of 
 cyber-attacks, cyber-criminals have continued to cause devastating financial losses to businesses,  enterprises, the governments, \textit{etc}. In 2018, the CSIS (Center for Strategic and International Studies) in partnership with McAfee has estimated the worldwide costs of cyber-attacks at about \$600 billion and it is predicted to cost the world \$6 trillion annually by 2021 \cite{Lewis2018CybercrimeReport}. Therefore, there is a need for more innovative techniques to assess and defend networked systems against cyber-attacks.

Offensive security testing techniques have been employed to assess the various security posture of networks by launching cyber attacks. Some of these testing techniques include: 1) the traditional penetration testing - where the testing focuses on identifying and exploiting the system and network vulnerabilities \cite{Goel:PenTest2015}, and 2) the red teaming (RT) - which assesses a network resilience against cyber-attack by emulating real cyber attackers \cite{Muehlberghuber:RedBlueTeam2013}. The RT moves beyond the penetration testing by imitating real steps that an attacker would necessarily take. 
However, conducting the red team exercise is a manual process and hence, the quality of security assessment can be depending on the quality of the red team members and their time and devotion to the test exercise.\\
On the other hand, automating the activities of the real attackers is faced with a great challenge of deciding the attacker's course of action.
Tools such as Attack Graphs (AG) \cite{Swiler:AG2001} have been used to represent possible sequences of actions that attackers may take to achieve the attack goal, but the AG focuses on analyzing the network vulnerabilities and producing a set of attack paths with no indication of the attacker's specific attack plan. Moreover, with the increasing size of modern networks, the AG has exponential complexity and thus causing scalability problems \cite{Ingols:AG2006}. Similarly, the Attack Trees (ATs) \cite{Schneier:AT1999} represents attacks as a tree with leaf nodes and child nodes, where leaf nodes show different ways of achieving the goal, and child nodes represent attack steps.
However, the ATs does not explicitly reflect the sequences of attack path nor specify a workable attacker's attack plan. \\
Hong and Kim \cite{Hong:HARM2012,Hong:MultiHaRM2016} addressed the scalability problem of AG by developing hierarchical models that combine (and separate the functionality of) the AGs and ATs unto two or more number of hierarchical layers (this model is named Hierarchical Attack Representation Models (HARM)).  The HARM mainly comprises of two layers: the upper layer which captures the network reachability information (using an AG that models only the reachability information) and the lower layer that captures the vulnerability information of each node in the network (using ATs). \\
Although the HARM has been used to generate the set of possible attack paths (similar to the AGs) to reach a target node, it has not been used to plan a rational attacker's possible attack action. Hence, more work is needed to strategically plan the attacker's and the defender's possible actions in the network.
Since the HARM is more scalable and adaptable compared to the AGs and ATs, we utilize its functionality to achieve this goal. Specifically, we develop a deterministic planning strategy (named metrics-based planning) with the HARM to systematically plan attacks for automated adversary actions. Moreover, we propose a novel framework named HARMer to automate the modeling and execution of cyber-attacks and threats detection. We carry out experiments in real network and Amazon Web Services (AWS) to demonstrate and validate the framework. The proposed framework will provide a way to automatically perform security analysis and evaluation of a real system by performing a red team and blue team operations. 

The major goals of this paper are summarized as follows.
\begin{itemize}
\item Develop a requirement specification for building automating cyber-attacks.
 \item Propose a framework for automating and assessing cyber-attacks activities.
 \item Develop an automated attack planner using a Graphical Security Model (GSM).
 \item Demonstrate the framework using a case study network and experiments on the AWS.
\end{itemize}

\textbf{Contribution highlight:}
It is difficult for network defenders to employ offensive testing techniques to evaluate a network security posture because they need to frequently search for a well-defined attack scenario that may be open to attackers. This process is time-consuming, costly, and impractical to perform regularly. Moreover, it depends on the quality of the team members to effectively plan and execute attacks. In this paper, we attempt to answer the following questions; (1) What is the approach that can be used to capture the attack scenario of a real attacker?, (2) How can real attacks be automated?, and (3) How long will it take to perform the automated attacks?

To answer these questions, we propose a novel framework for automating the modeling of cyber-attacks. The framework will support the automatic assessment of network security by collecting attack information and then exploiting them, just like a real attacker would necessarily perform. By doing so, a defender can understand the appropriate network weak spots and deploy the best form of available cyber defense. \\
Existing frameworks that used the AGs to identify overall potential attack paths suffer from computational complexity. As a result, it is challenging to represent a full range of cyber-attacks with the AGs due to the numerous possibilities and choices that are available to the attacker. In this paper, we incorporate a scalable security model (HARM) to reduces this complexity \cite{Hong:MultiHaRM2016}.
Moreover, we develop and automate three new metric-based attack planning strategies that automatically generate a more specific and realistic attack path to use (because the HARM or AGs does not explicitly specify what attack path will be exploited per time). In addition, we model the networks with nodes and edges, in which the nodes have various attributes that model the node components such as the operating system (OS), vulnerabilities, open port, \textit{etc} in order to allow for multiple simulations in different scenarios. In Table \ref{tbl:contribution}, we highlight our contributions compared to similar approaches. We use the symbols \checkmark and \ding{55} to show paper contribution and those that did not, respectively.

The rest of this paper is organized as follows. Section~\ref{sec:related_work} gives summary of the related work. Section~\ref{sec:methodology} discusses methodology, requirement analysis, and the automation framework. Section~\ref{sec:planning_strategy} describes the proposed attack planning strategies. Section~\ref{sec:framework_demo} presents the illustration of the attack framework using a case study. In Section~\ref{sec:Experiments}, we present our experiments and results based on Amazon's AWS using two network models.  Section~\ref{sec:discussion} discusses our results, limitations and future work.  Lastly, Section~\ref{sec:conclusion} concludes the paper.

\begin{table*}[!ht]
 \scriptsize
\centering
\caption{Contributions highlight}
    \label{tbl:contribution}
\begin{tabular}{lccccccccc} \hline 
 & \cite{kotenko2006attack,wang2008attack,Poolsappasit:BAG2012} & \cite{moskal2016knowledge} & \cite{bergin2015cyber} & \cite{NOOR2019467} &\cite{yuen2015visual} & \cite{Obes2010:attackplanning,Elsbroek2011:Fidius,Sarraute2011:Algorithmplanning} & \cite{iqbal2020scerm}&\cite{Hong:HARM2012,Hong:MultiHaRM2016}& This paper \\ \hline \hline
Automation framework & \ding{55}  &\checkmark  &\checkmark &\checkmark &\checkmark & \ding{55} &\checkmark & \ding{55} &\checkmark \\ 
Detailed attack planning & \ding{55}  &\ding{55}  &\ding{55} &\checkmark &\ding{55} &\checkmark & \ding{55} & \ding{55} &\checkmark \\ 
Scalable GSM &\ding{55}   & \ding{55}   & \ding{55}  & \ding{55} & \ding{55} &\ding{55} & \ding{55} &\checkmark &\checkmark \\ 
Experiments &\checkmark   &\checkmark  &\checkmark  &\checkmark & \checkmark &\ding{55} & \checkmark & \ding{55} &\checkmark \\ \hline
\end{tabular}
\end{table*}

\section{Background and Related Work}
\label{sec:related_work}
We discuss the state-of-the art work on automating cyber-attacks and defenses.

\subsection{Security Model Automation for Red Team and Blue Team}
\label{sec_sub:relatedWOrk_model_attack_Def}
There are a lot of works that addressed the problem of assessing the security of network systems using different types of automation approaches. We discuss the related work in two aspects: security models, and attack \& defense framework.

\textbf{Security Models:}
One of the popular use of automation for red team activities is the use of AGs. The AG provides a way for the red team to generate possible sequences of attack steps to gain access to a target using network reachability information and a set of vulnerabilities. The work of Phillipsi \& Swiler \cite{Phillipsi:VulAnalysis} is one of the earlier work that developed a graph-based tool to assess the risks to a networked system by identifying the set of attack paths with a high probability of success or low attack costs for the attacker. This tool provided a way to test the effectiveness of defenses (such as intrusion detection systems, firewall rules changes, etc). 

Sheyner \textit{et al.} \cite{sheyner2002automated} presented an automated approach to generating and analyzing AGs based on symbolic model checking algorithm. Besides, they performed minimization analysis on the AG to determine the minimal sets of atomic attacks that must be prevented in order to guarantee that the attacker cannot reach his goal. Kotenko and Stepashkin \cite{kotenko2006attack} utilized the AGs to simulate and evaluate the attacker's actions (based on vulnerabilities). To improve security, Kotenko and Stepashkin checked the various properties of the AGs and then used various security metrics to determine ways to prevent possible attacks. 

Wang \textit{et al.} \cite{wang2008attack} proposed an AG-based probabilistic metric to measure the likelihood of sophisticated attacks combining multiple vulnerabilities to reach the attacker target. Poolsappasit \textit{et al.} \cite{Poolsappasit:BAG2012} proposed Bayesian AGs to quantify the likelihood of a network being compromised at different levels. Based on the level's information, Poolsappasit \textit{et al.} developed a security mitigation and management plan for the network administrator. \\
\textit{Summary: }The aforementioned AG approaches focused on generating a set of attack paths to the attack goal with no indication of a specific attack path that at adversary may use per time. Hence, it is difficult to use the AGs to automate the real-world interaction between the attacker and the defender since no specific plan is shown.\\

\textbf{Framework for cyber-attacks:}
An attack framework will provide a structure and flow to combine the analysis and evaluations of cyber-threats . In this section, we present the state-of-the-art automation framework.

Moskal \cite{moskal2016knowledge} presented a framework for modeling cyber-attack behaviors for use with existing attack simulators in order to analyze the effects of single or multiple attackers on a network. This framework utilizes Cyber Kill Chain behavior to model an attacker's decisions while taking into account what the attacker knows, how the attacker learns about the network, the vulnerabilities, and targets.
Similar to our work is the extension provided by Moskal \textit{et al.} \cite{moskal2018cyber}, which proposed the red and blue team's simulation framework to show the interplay between an attacker and defender. The framework was defined based on the network, the attackers, and the intentions, the dependencies between the attacker and the network including capabilities and preferences. Furthermore, they showed an assessment approach of how different attack scenarios may occur under different attacker's intent, opportunity, capability, and preference against a network configuration. However, our work is richer in terms of attack planning strategies.

Matherly \cite{matherly2013red} provided a theoretical framework to investigate and identify the best strategy for combining red teams and social psychology techniques to improve adversary prediction. Bergin \cite{bergin2015cyber} presented a cyber-attack and defense simulation framework to support the modeling and simulation of cyber-attack and defense for training and assessment. The work focused on modeling and simulation for cybersecurity of autonomous vehicle systems (wireless communications) used by US Armed Forces.
Applebaum \textit{et al.} \cite{applebaum2016intelligent} developed a framework that used MITRE framework-Adversarial Tactics, Techniques, and Common Knowledge (ATT\&CK). Their framework specifically takes into account the post-compromise effect that an adversary can take in a network.

Choo \textit{et al.} \cite{choo2007automated} leveraged parallel processing,
evolutionary algorithms and agent-based simulations to develop an automated RT (ART) framework for a military operation. The framework consists of (1)  ART parameter setting interface which will allow the initial selection of the parameters that are to be varied, (2) ART controller - controls and coordinates the whole process of the framework, (3) the simulation model-dependent modules add a layer of data flow to and from the framework and simulation model of the parameters to be executed, (4) the EA module prepared the parameters for the simulations and analysis using any of the EAs, (5) the condor provides a job queuing, scheduling policy and resource management for distributed computing and (6) the output module is used to provide feedback, update and run results. In Chua \textit{et al.} \cite{chua2008automated}, the capability of the ART framework in  \cite{choo2007automated} was evaluated against the manual RT using two maritime security scenarios.

Yuen \textit{et al.} \cite{yuen2015visual} developed an ART framework that uses automated planning and knowledge representation techniques to conduct the RT exercise. The high-level view of the framework consists of the world model (i.e., the overall system that is being red-teamed), AI planner, AG generator, threat analysis, course of action planning, change deployment, \textit{etc}.

Noor \textit{et al.} \cite{Noor2019:MLFramework} presented a machine learning framework for investigating data breaches based on common patterns from threat repositories. The framework reasons on security incidence by mapping low-level threat artifacts to high-level adversary tactics, techniques, and procedures in a way that machines can identify these connections with certain probabilities. In \cite{ghazi2018supervised}, the authors presented a machine learning-based approach to automatically extract cyber threats information such as attack patterns and techniques that may represent attacker behaviors or attack exploits. \\
\textit{Summary:} These approaches are different from our framework as they have focused on mapping information from existing repositories or threat artifacts based on the probability of attacks while our proposed automation framework is based on real-time attack information and execution on a network. Moreover, the frameworks lack sufficient attack planning methods or they are based on a theoretical framework.

\subsection{Attack Planning}
\label{sec_sub:relatedWOrk_planning}

Identifying a workable attack path can be time-consuming for the RT, and so automated planning techniques are being considered as a feasible method of discovering possible attack paths for automating the RT agent.
There are a few works on the application of planning techniques for reasoning in emulation/simulation of attacker behavior. Boddy \textit{et al.} \cite{Boddy2005:courseAction} presented an approach for the generation of adversary courses of action from the initial state to the target machine using a classical planning technique. This planning approach was used to predict the attacker's actions.
Obes \textit{et al.} \cite{Obes2010:attackplanning} used Planning Domain Definition Language (PDDL) description of network hosts, vulnerabilities, and exploit to generate attack paths which were integrated into a penetration testing (pentest) tool. Elsbroek \textit{et al.} \cite{Elsbroek2011:Fidius} also used the PDDL to generate attack paths for a pentest tool.

Sarraute \textit{et al.} \cite{Sarraute2011:Algorithmplanning} addressed the problem of attack planning by taking into account uncertainty about the results of the attacker's actions, then modeling it as the probability of success for each action. In another work, Sarraute \textit{et al.} \cite{Sarraute2011:penetrationPOMDP} modeled the attack planning problem in terms of Partially Observable Markov Decision Processes (POMDP) for a pentest. Applebaum \textit{et al.} \cite{applebaum2016intelligent} and Miller \textit{et al.} \cite{miller2018automated} used classical planning, Markov Decision Processes, and Monte Carlo simulations to plan attacks for an automated red teaming system (named Caldera). 

Ghost \textit{et al.} \cite{Ghosh2009:intelligentAG} proposed an approach based on a search algorithm for the AG that automatically generates attack paths (i.e., using a planner as a low-level module). Durkota \textit{et al.} \cite{Durkota2019hardening} used AGs to determine attacker's next actions. The authors compute the attacker's set of possible actions based on AG reduction. 

Randhawa \textit{et al. } \cite{randhawa2018mission} presented an automated planning and cyber red-teaming system called  Trogdor. Randhawa \textit{et al. } described Trogdor as a mission-centric red-teaming and defensive decision support system that can generate and visualize potential attack paths for known vulnerabilities for a networked system.  The Trogdor used domain ontologies to describe the target environment; the network information and inter-dependencies between them, and the known software or hardware vulnerabilities.
Ghanem \& Chen \cite{ghanem2020reinforcement} proposed a reinforcement learning (RL) technique, where the system  (named IAPTS) is modeled as a POMDP, and tested using an external POMDP-solver with different algorithms. According to Ghenem \& Chen, the proposed system can act as a module and can be integrated with most of the industrial pentesting frameworks to improve efficiency and accuracy. Similarly, Zennaro and Erdodi \cite{zennaro2020modeling} presented a penetration testing approach using different RL techniques in a simulation. The focus of their work is to understand the feasibility of using RL techniques for RT. In our work, we focused on automatic attack execution on a real network.  

Technologies such as Parallel Computing and Evolutionary Algorithms (EAs) are used to plan the red teaming exercise as well, where the Parallel Computing is leveraged to perform millions of simulations runs in an automated way, while EAs is used to optimize the required fitness value that can serve as the objective function. Specifically, the evolutionary algorithm is used to plan and decide defense options within the least amount of time. For example, Choo \textit{et al.} \cite{choo2007automated} used the evolution algorithms as the search algorithm to search for red parameters that result in the ``defeat” of blue then fix the parameter.\\
\textit{Summary:} Several complex attack planning strategies have been proposed, however, only a few have been used for automating the RT agent, while others have been used in isolation (without a defined framework) and not in a real network environment.


\section{Methodology of the Automation of Threat and Defense Modeling}
\label{sec:methodology}

Automating the attacker's and defender's operations will require detailed modeling of the interaction between the attacker and the defender on a target network. To achieve this, we develop a framework for the automation of threats, attacks, and defenses. First, we specify the requirements for the automation in Section \ref{sec:requirement_analysis}.
We use high-level cyber-attacks descriptions and then use an attack model similar to the cyber kill chain and the MITRE ATT\&CK for our attacker model. We believe using an accepted security model will provide a realistic result hence, we adopt a deterministic planning strategy (metric-based) with the HARM to model the attacker's course of actions with the network components, respectively. 

\subsection{Requirements Analysis on the Automation of Threat and Attack Modeling}
\label{sec:requirement_analysis}

In this section, we discuss the requirements for the automation of cyber-attacks. Here, the main focus is to provide the detailed requirements analysis to model the attacker along with his actions for the red team model. These requirements analysis steps will support the simulations and executions of attacks and defenses for the real systems. There are four steps in the requirements analysis and each step will correspondingly describe the requirements for each phase in the automation framework (shown in section \ref{sec:framework}). We list the requirement analysis steps as follows; (i) Data collection, (ii) Graphical Security Model (GSM) Construction, (iii), Attack Planning, and (iv) Attack Execution and Evaluation. We use Figure \ref{fig:proposed_framework} to explain these steps as follows.

The data collection phase requires the network, threat, and attacker information, and defenses information from the real environment, where the network consists of hosts, the host's configurations, vulnerability, open and close ports, service, \textit{etc}. These network information can be collected using existing tools such as NMap \cite{Nmap} and OpenVAS \cite{OpenVAS:scanner}. Similarly, existing databases of attack tactics and techniques can be used to gather information about attacks and their behavior. For instance, the MITRE ATT\&CK framework \cite{Mitre_framework} can be used to provide atomic actions on real-world attacker's tactics, techniques, and common knowledge. Besides, the security administrator can provide other information regarding the attacker's knowledge and the list of available hardening options. 

In the GSM construction phase, a GSM and metric(s) are required to analyze and process the data collected. Specifically, the GSM uses the network information (hosts, vulnerabilities, services, ports, \textit{etc}) to construct and build a model that represents the security posture of the network at that time, and security metric to quantitatively measure the security posture. Using the GSM and a specified security metric, attack paths and metric values can be calculated and an output will be generated. 

The attack planning phase requires a planning strategy and an attack language. The planning strategy determines the attack plan while the attack language will bridge the attack plan with the attack execution (i.e., next phase) in a universal way. 
As part of this phase, the output generated from the data analysis phase is passed to the planning strategy and the attack language to generate and translate the attack plan to be executed, respectively.
In the attack execution and evaluation phase, attack tools (such as Metasploit \cite{Metasploit} or any other attack tools) is required. Also, an attack language is needed for the attack execution since it needs to translate the attack plans before executing them. 
This phase recursively requires input from the attack planner to determine viable attack paths for execution. Attack paths that cannot be executed are eliminated and then the output from the planner is updated for the next execution. The execution is reviewed automatically and the results are used for the next iteration(s). 
\subsection{An Attack Automation Framework}
\label{sec:framework}

\Figure[h!][width=0.8\textwidth]{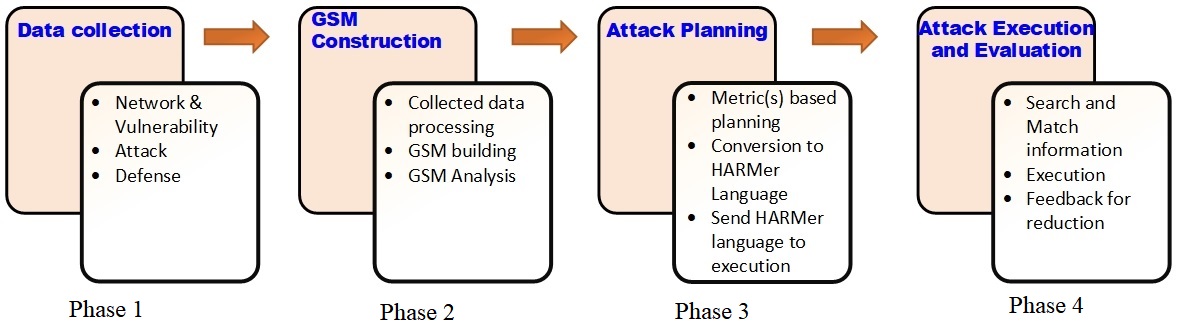}
{A framework for attack automation\label{fig:proposed_framework}}

The main goal of the framework is to develop a method to automatically exploit the vulnerabilities of a network system. The end goal of our work is to evaluate the effectiveness of defense strategies against cyber-attacks in real systems by lunching real attacks. However, in this paper, we focused on only \textit{automatic attack execution}, while the defense automation is left for future work. In Figure \ref{fig:proposed_framework}, we show the different phases of the attack automation framework and we list them as follows; 1) Information collection, 2) Security model construction and analysis, 3) Attack planning, and 4) Attack execution and evaluation. We explain each phase as follows.

\subsubsection{Data collection}
\label{sec:info_collection}
A lot of information can be collected from a network environment \cite{ZHOU20189}. However, this framework uses only the relevant information required to construct the attack model. The proposed framework incorporate vulnerability scanning tool, and network and open ports discovery tool together to automatically collect information. 
However, the framework is not limited to the information collected via the scanning and discovery tools, as network administrators are allowed to provide other information as well. For example, a network administrator may have a network map of the devices found on his network including their configuration information (an example network map is shown in \cite{Mansfield:networkMap}). Such a map (which provides the network topology) can be used as input in this phase. Besides, security tools such as OpenVAS \cite{OpenVAS:scanner} Nessus \cite{Nessus} and Nmap \cite{Nmap} can be used as well to collect the hosts' vulnerability information, operating systems, services port, \textit{etc}.  

\subsubsection{Security model construction and analysis}
\label{sec:security_model_analysis}
In the second phase, a two-layered HARM of the network is generated using the information that is collected in phase 1. All potential attack paths are captured and enumerated in the HARM, whence the possible attack scenarios are well captured. For security analysis, the security decision-maker can select the security metrics to use with the security model. Here, the computed security metrics via the model will be used to make decisions on the attack plan. Therefore, this phase evaluates every attack path based on their risk, damages, or probability of a successful attack, depending on the selected security metrics.

\subsubsection{Attack Planning}
\label{sec:attack_plan}

This phase is responsible for planning and generating actions for the adversary (attacker) agent. Here, a plan may include a response from the next host/target, port scanning and IP ranging, or targeted actions such as exploiting a software vulnerability of a host, sending a spear-phishing email, \textit{etc}. Various approaches may be used with the HARM to strategically generate possible attack plans. We describe these approaches in Section \ref{sec:planning_strategy}.\\
We have chosen to use metrics-based attack planning and HARM to generate attack plans for attackers in the form of attack scenarios. We can generate deterministic attack plans based on the metrics used in the HARM. The attack plans can be formulated in the attack language.  
The main reason to use an attack language is to allow the created attacks plans to be universal and useful for different types of attacks and defense tools. Moreover, attack plans written in a universal attack language can be converted to a suitable format used for such attacks and defense tools.\\

\subsubsection{Attack Execution and Evaluation}
\label{sec:attack_execution}
 
\begin{figure}[h]
\centering
\includegraphics[width=0.2\textwidth]{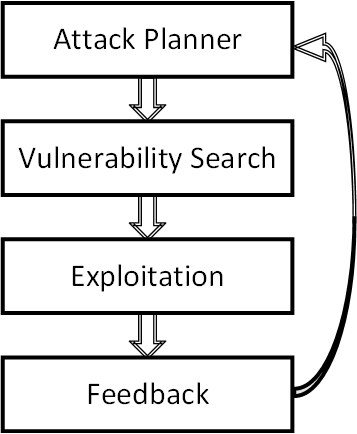}
\caption{Attack Planner with Attack Execution Flow} 
\label{fig:planAndExecution_flow}
\centering
\end{figure}

 In this paper, we initially choose to use Metasploit as an attacking tool for the exploitation of vulnerabilities. Here, the output generated from the attack planning phase is fed to the attack execution and evaluation phase as inputs for executions.
 Our current implementation is not supporting conversion from the security model language to the Metasploit compatible outputs. Instead, we implemented a direct conversion from metrics-based planning outputs to a Metasploit compatible format. The attacks are performed on the vulnerabilities that have been discovered from the initial `information collection' phase.
 It is important to highlight that the attack execution phase recursively works with the attack planning phase. \\
In Figure \ref{fig:planAndExecution_flow}, we describe the relationship between the attack planning phase with the attack execution and evaluation phase. We describe each of the stages as follows.

\begin{itemize}
    \item \textit{Attack Planner} carries the list of the attack paths (passed from phase 2), each path is an ordered list of nodes (representing the hosts) and each node contains hosts information and vulnerability information. Besides, it also carries the attack plan to be executed.
    
    \item \textit{Vulnerability Search} will extract the keywords (e.g.Common Vulnerabilities and Exposures (CVE), OS) which describes possible vulnerability information of each node and return an exploit module (e.g.exploit/windows/smb/ms17\_010\_psexec) that can be understood by the adversary program.
    \item \textit{Exploitation} will extract the host information of each node, along with the exploit module extracted by the Search, then start the exploitation process.
    \item \textit{Feedback} will be generated after the sequence of exploitation is finished. Specifically, feedback is generated for the following two scenarios; (1) all the nodes on the attack path have been successfully traversed and exploited. (2)  some (or all) of the nodes on the attack path cannot be exploited (which terminates' the process of the attack for that particular path). In this case, an attack will fail for different reasons such as a node on the attack path is not exploitable, the vulnerability information provided by the planner (phase 3) does not match with the host's real vulnerabilities,  or the host is not down/not available.

\end{itemize}

\section{A Propose Attack Planning Strategy}
\label{sec:planning_strategy}
Reasoning and planning an adversary course of action to achieve a target is a difficult task. An automated planner can logically decide the sequence of actions to achieve a set of goals. However, depending on the adversary's learning ability of the network, attack plans may be conceived in various ways. For example, an attacker that has global knowledge of the network environment can plan varying attacks compared to the attacker with partial knowledge. 
Using the HARM, various possible attack plans can be systematically generated for the red team exercise. In this regard, we consider attackers with the following network knowledge with the HARM; (1) Global learning, and (2) Partial learning. We propose the following metric-based attack planning strategies. We consider that these planning strategies may be used with other automated frameworks that use GSM (such as AGs)). 

\begin{itemize}
  \item Path-based approach: for example, the shortest attack path
 \item Composite metrics (e.g., probability of attack success with attack paths)
 \item     Atomic metric (e.g., attack cost only)
\end{itemize}

In the next sections, we present and describe algorithms for computing the attacker plan using three different metric-based approaches. These algorithms are the key components in the attack planning phase of the automated framework. In Table \ref{tbl_notations}, we provide the notations used for the rest of the paper.

\begin{table}[h]
\caption{Notations}
\label{tbl_notations}
\begin{tabular}{l|l}
\hline
Notation & Meaning \\ \hline
$h$ & is a host \\ \hline
$ap$ & is an attack path \\ \hline
$AP$ & is a set of attack paths \\ \hline
$p_{h}$ &  is the probability of attack success of $h$ \\ \hline
$aim_{h}$ & is the attack impact $h$ \\ \hline
$r_{h}$ & is the risk value of a host \\ \hline
$aim_{ap}$ & is the attack impact of the path $ap$ \\ \hline
$AIM$ & is the network level attack impact value\\ \hline
$SP$ & is the shortest attack path metric \\ \hline
$H$ & is set of hosts \\ \hline
$E$ & is set of edges \\ \hline
\end{tabular}
\end{table}

\subsection{Path-based approach}
\label{path_based_planning}
Attack vectors enable attackers to exploit system vulnerabilities and to reach their goals. One aspect of an attack vector is an attack path. An attack path is a sequence of steps with one or more vulnerabilities that can be exploited by attackers to gain access to specific assets in a network, hence, forming an exploitable attack path between the assets. In this section, we consider the attack paths to develop an attack planning strategy. \\
Assuming that an attacker has global knowledge of the target network, where the attacker knows the network topology, assets vulnerability information, and the specific target asset. Then a rational attacker is more likely to select and compromise the set of machines with a shorter distance to the target. Here, we can utilize one or more of the path-based security metrics to strategically plan an attacker's sequence of actions to accomplish the attack goal. 
We use the shortest path metric \cite{Phillipsi:VulAnalysis} to generate deterministic attack plans based on analysis on the HARM. We define the shortest path metric as follows.\\

\textbf{Shortest path metric:} The shortest attack path is defined as the minimum distance from the attacker to the attacker's goal and it is formally defined by Equation \eqref{eq:sp}. This metric equation finds the smallest number of sequence of hosts to the target machine that an attacker must use to achieve the attack goal. 
The HARM can model the security posture of a network, and so can be used to compute attack paths and risks associated with host vulnerabilities based on the Common Vulnerability Scoring System (CVSS) Base Score (BS) metrics. Specifically, we use the upper layer of HARM to find the potential attack paths to the attacker's goal. Hence, this provides us with the set of the potential paths that will be open to the attacker.\\
The shortest attack path metric is calculated in the upper layer of HARM using the formula in Equation \eqref{eq:sp}, where $\mathbb{AP}$ is a set containing different attack paths.

\begin{equation} \label{eq:sp}
SP =\mathop {min}{|\mathbb{AP}|}
\end{equation}

\begin{algorithm}[h]
    \caption{Attack plan using Shortest Path Metric}
    \label{alg:shortest_path_planning}
    \begin{algorithmic}[1]
        \Procedure{Find Shortest Attack paths(\textit{Network})}{}
         \State{Initialise \textit{Paths} $\rightarrow$ [ ]}
         \State{Initialise \textit{Full Plan} $\rightarrow$ [ ]}
         \State{\textit{HARM = HARM Model(Network)}}
         \State{\textit{Paths} $\rightarrow$ \textit{Find All Paths(HARM)}}
         \State{\textit{SP} = \textit{min(Paths)}}
        \ForAll{\textit{path} $\in$ \textit{Paths}}
        \If{\textit{|path|} $==$ \textit{SP}}
        \State{append \textit{path} to \textit{Full Plan}}
        \EndIf
        \EndFor
        
        \If{\textit{|Full Plan|} $==$ \textit{1}}
          \State{$max\_risk =$ 0}
          \State{Initialise \textit{Reduce Plan} $\rightarrow$ [ ] }
           \State{\textit{Plan} $\rightarrow$ $\varnothing$}
        \ForAll{\textit{path $\in$ Full Plan}}
        \State{\textit{Risk} $=$ $0$}
        \ForAll{\textit{host $\in$ path}}
        \State{\textit{Risk} $=$ $p_{host} \times aI_{host}$}
        
        \EndFor
       
        \If{\textit{Risk} $> max\_risk$}
        \State{$max\_risk$ $\rightarrow$  \textit{Risk}}
        \EndIf
         \State{append \textit{path} to  \textit{Plan}}
          \State{\textit{Plan} $\rightarrow$ \textit{path}}
        
        \EndFor
        \State{\textit{Reduce Plan} $\rightarrow$ \textit{Plan} }
        \State{return \textit{Reduce Plan}}
       
       \Else
        \State{return \textit{Full Plan}}
        \EndIf
        \EndProcedure
        
    \end{algorithmic}
\end{algorithm}

Computing the shortest attack path metric may generate a set of possible attack scenarios consisting of multiple attack paths for the attack execution (phase 4) \cite{Enoch:evaluation2018}. However, a single attacker would typically choose a single attack path per time. Hence, a suitable approach is needed to carefully select the most appropriate attack plan to execute first given multiple attack paths. To address this, we rank the attack paths generated. Specifically, we apply prioritization to the attack paths based on risk, where the most concerning or the most critical attack path is selected first. Other criteria such as the most likely attack to succeed, the attacks with high impact if successful, the expensive attacks (in terms of economic metrics), \textit{etc} can be used as well. To prioritize, we compute the risk (i.e., the expected value of impact) associated with each attack path based on CVSS BS. The formula for the path-based risk is given by Equation \eqref{eq_risk_ap}.

\begin{equation} \label{eq_risk_ap}
\begin{array}{ll}
Risk_{ap}=\sum\limits_{h \in ap}{p_h \times aim_h}, & ap\in \mathbb{AP}
\end{array}
\end{equation}

Algorithm \ref{alg:shortest_path_planning} describe the shortest attack path metric planning. We summarise the algorithm as follows. A network and set of security information are provided as input to the algorithm. HARM is used to calculate possible attack scenarios (attack paths) to a specific target. Equation \eqref{eq:sp} and Equation \eqref{eq_risk_ap} are used to compute the set of shortest attack paths and the attack paths prioritization (i.e., if there are multiple shortest paths) via the HARM, respectively. Then, the final plan is returned as the attack plan to be used in the third phase of the automation framework.

\subsection{Composite metrics approach}
\label{composite_metrics_planning}
Often, attackers' may be having multiple interests, which individual security metrics may not be able to capture. For example, an attacker may be interested in performing a multistage attack, and also be interested in compromising the hosts with a high likelihood of attack success to reach the target.
To capture such scenarios, we propose an attack planning strategy based on composite metrics. Composite metrics combine individual metrics to form a new metric. For example, the attributes of a multistage attack (i.e., attack paths) are combined with the probability of attack success metric to form a metric named probability of attack success on paths \cite{Enoch:Composite2017}. This metric assesses the likelihood of an attack to be successful via an attack path and it is calculated by equation \eqref{eq_asp_ap} and \eqref{eq_pr}. The attack path with the maximum \textit{P} (from \eqref{eq_pr}) is extracted for the attack planner. 

The procedure for this approach is similar to Algorithm \ref{alg:shortest_path_planning} but with a different type of metric. As a result, we did not show an algorithm for it.

\begin{equation} \label{eq_asp_ap}
\begin{array}{ll}
p_{ap}=\displaystyle \prod\limits_{h \in ap}{p_h}, & ap\in \mathbb{AP}
\end{array}
\end{equation}

\begin{equation} \label{eq_pr}
P=\mathop {max}\limits_{ap \in AP}{p_{ap}}
\end{equation}

\subsection{Atomic metric approach}
\label{sec_sub:atomic_metrics_planning}

In the previous subsection, we have considered an attacker with full knowledge of the network, where the attacker has perfect information about the network. However, a real attacker may not know the complete network reachability information (i.e., the topology), but she/he may have partial knowledge about a few networked hosts that are available to the public.
Hence, we assume an attacker with incremental learning ability and a specific target machine as the attack goal. Then, we develop a planning strategy base on atomic security metric.\\
One important decision factor to both an attacker and the defender is costs \cite{Enoch:Economicmetrics2018}. We consider the cost of attacks based on attack effort or the difficulty of exploiting the vulnerabilities (i.e, from the attacker's perspective). The attack cost of exploiting host vulnerabilities can be estimated using different ways. For instance, the CVSS \cite{Schiffman2004cvss} provides a vulnerability exploitability score which shows the difficulty of exploiting the vulnerability. We can use the CVSS vulnerability exploitability score as an attack cost metric for vulnerabilities. The hosts' attack costs metric is calculated from the lower layer of HARM based on the host AT. A detailed explanation of the calculations of the attack costs metrics based on ATs is provided in \cite{Enoch:Composite2017}.

We use the attack cost metric (from the perspective of the attacker) to determine the attacker's choice of the host to exploits. Typically, an attacker may choose hosts with a lower attack effort for a multistage attack on a target machine. We use this idea to develop an attack planning strategy when the attacker has only partial knowledge of the network. The attacker moves incrementally from the initial host to an adjacent host based on their reachability and attack cost value until the attack goal is reached.

Algorithm \ref{alg:incremental} is used to explain this attack planning approach with the following context. The attacker has access to (and knows) only the device that is open to the public (e.g., a web server). the attacker will be able to gain root control of the devices open to the public, leveraging an existing vulnerability. This could happen because the vulnerability couldn't be patched due to software dependencies or time to patch. The attacker may be able to discover the hosts that are reachable from the web servers (e.g., the application server) that are located in the internal network. Then, the attacker can scan the hosts for vulnerabilities and then exploit the easiest vulnerability (low attack effort) then move to the next host. Here, the firewalls may not be able to block the connection as the attacker is using a legitimate user account and privileges. The target only accepts connections from the internal networked hosts, and so the attacker can reach the target and then escalate the target' privilege to admin privilege using existing vulnerabilities. \\ \\

\begin{algorithm}[h!]

    \caption{Attack planning with incremental learning using attack cost metric}
    \label{alg:incremental}
    \begin{algorithmic}[1]
        \Procedure{Incremental learning(\textit{Network})}{}
         \State{Initialise \textit{Path} $\rightarrow$ [ ]}
         \State{\textit{HARM} $=$ $(H, E)$}
          \State{\textit{K} is the set of attacker's entry points and initial location $K \subseteq H$}
      
        \ForAll {$K \subseteq H$}
            \State{\textit{Scan for vulnerabilities} }
            \State{\textit{min ac} = 10}
            
            \State{\textit{Plan} $=$ $\varnothing$}
            \ForAll{\textit{host $\in$ K}}
            \State{compute $ac_{host}$}
            \If{\textit{$ac_{host}$} $\leqslant$  \textit{min ac}}
            \State{\textit{min ac} $=$  $ac_{host}$}
              \EndIf
             \State{\textit{Plan} $\leftarrow$ \textit{host}}
            
             \EndFor
             
            \State{Append \textit{Plan} to \textit{Path} }
        \EndFor
        \While{\textit{attack target} $\notin$  \textit{Path}} 
         \State{\textit{lastElement}=  Get Last element  \textit{(Path)}}
            \State{\textit{SetHosts} $\leftarrow$ Get hosts adjacent to \textit{lastElement} }
             \State{\textit{Next Plan} $=$ $\varnothing$}
              \State{\textit{min ac} = 10} 
            \ForAll{\textit{host $\in$ \textit{SetHosts}}}
            \State{compute $ac_{host}$}
            \If{\textit{$ac_{host}$} $\leqslant$  \textit{min ac}}
            \State{\textit{min ac} $=$  $ac_{host}$}
              \EndIf
             \State{\textit{Next Plan} $\leftarrow$ \textit{host}}
            
             \EndFor
             
            \State{Append \textit{Next Plan} to \textit{Path} }
        \EndWhile
        \EndProcedure
    \end{algorithmic}
\end{algorithm}

\section{Attack Modeling and Automation with the Proposed Framework}
\label{sec:framework_demo}
We perform an experiment to illustrate our proposed framework. 
We focused on analyzing network security and demonstrating each phase of the framework using a case study network and simulations. The metrics, case study network, and the assumptions used in the experiments are described in the next section. 

\subsection{Metrics}
\label{metrics} 
We briefly describe the metrics build into the framework to evaluate the security posture. We used these metrics to show the final security evaluation. More details about the metrics are provided in \cite{cheng2014metrics, Enoch:evaluation2018, Cremonini:ROA2005}.
\begin{itemize}
    \item \textbf{Number of attack scenarios (NAS)}: This metric calculates the total number of possible attack scenarios with known exploit modules \cite{Enoch:evaluation2018}. Since our framework is based on known attacks and attack exploits, all possible attacks generated can be exploited. However, these metrics returns the total number of ways the attacker can compromise the target system. We calculate this metric using the formula $NAS(HARM)= \mathbb{|AP|}$.
    
    \item  \textbf{Percentage of severity level of vulnerabilities} (attacks): These metrics are incorporated from Cheng \textit{et al.} \cite{cheng2014metrics}. Severity level of vulnerabilities indicates the critical nature of the possible attacks found on the network hosts and it is calculated as a percentage with respect to all the exploitable vulnerability found. The metrics are grouped based on CVSS base score into \textit{high} (High), \textit{medium} (Medium), and \textit{low} (Low), where high is base score 7.0 - 10.0, medium - 4 - 6.9 and low is 0.0 - 3.9.
    
     \item \textbf{Return on the attack (ROA)}: This metric calculates the benefit an attacker gain when they are able to exploit vulnerabilities. It is a metric from the view of the defender \cite{Cremonini:ROA2005}. We calculate the host risk metric based on CVSS BS and the attack costs are assigned based on severity of vulnerability metric (where a vulnerability with severe metric value is assigned low costs to show that it requires less effort to exploit the vulnerability). For example, the CVSS \cite{Schiffman2004cvss} provides a vulnerability exploitability score which shows the difficulty of exploiting the vulnerability. We calculate the metric using equation \eqref{eq_roa1} and \eqref{eq_roa2} at the path level and network level, respectively. More details is provided in \cite{Enoch:evaluation2018}.
    \begin{equation} \label{eq_roa1}
    \begin{array}{ll}
    roa_{ap}=\sum\limits{\frac{r_h}{ac_h}}, & h\in ap
    \end{array}
    \end{equation}
    
    \begin{equation} \label{eq_roa2}
    \begin{array}{ll}
    ROA=\sum\limits {roa_{ap}}, & ap\in \mathbb{AP}
    \end{array}
    \end{equation}
    
    \item \textbf{Attack impact}: measures the potential harm caused by an attacker to exploit a vulnerability. We denote this metric as $AIM$ and calculate it using Equation (\ref{path_level}) and equation (\ref{network_level}) \cite{Enoch:evaluation2018}. We calculate the impact metric based on CVSS metrics \cite{CVSS}. The attack impact metrics indicate the level of damages associated with attacks and possible attack paths. Here, the security manager is expected to reduce the impact values.

    \begin{equation} \label{path_level}
    \begin{array}{ll}
    aim_{ap}=\sum\limits {aim_h}, & h\in ap
    \end{array}
    \end{equation}
    
    \begin{equation} \label{network_level}
    \begin{array}{ll}
    AIM=\sum\limits {aim_{ap}}, & ap\in \mathbb{AP}
    \end{array}
    \end{equation}
\end{itemize}

\subsection{Case Study}
\label{sec:sub_case-study}

Many enterprise networks only realized that they have been attacked after discovering disparity in their activities or log files \cite{Burton:Networkmgt2003}. Businesses and enterprise networks' administrators must understand their security posture to provide an optimum defense. In this section, for simplicity, we use a small corporate network as our case study. The purpose of this section is to demonstrate the applicability and usability of each phase of the framework.  \\

\textbf{Network Model}\\
A small subset of an operational university network is used as our network. The network consists of 7 hosts with each host running windows operating system (OS).  
We assume that the network has one firewall which controls access between the networked hosts, and the firewall rules for the network are shown in Table \ref{tbl:firewall}. In this paper, for confidentiality, we did not use the actual IP addresses that were collected from the real network directly. However, we use the hosts' OSes and vulnerabilities as collected. \\
We assume one of the hosts with IP address 206.171.47.7 contains sensitive financial data and it is protected by the firewall, such that there is no direct access to the host - 206.171.47.7 (e.g., in a 3-tier network architecture). However, the network users are able to reach the hosts 206.171.47.7 after passing through other hosts e.g., 206.171.47.1 or 206.171.47.2. \\

\begin{table} 
\caption{The network firewall rules}
    \label{tbl:firewall}
\begin{tabular}{l}
\hline
Host Rechability \\ \hline \hline
Attacker $\rightarrow$ 206.171.47.1, 206.171.47.2 \\ \hline
206.171.47.1 $\rightarrow$ 206.171.47.3, 206.171.47.4, 206.171.47.7 \\ \hline
206.171.47.3 $\rightarrow$ 206.171.47.5, 206.171.47.7 \\ \hline
206.171.47.4 $\rightarrow$ 206.171.47.6 \\ \hline
206.171.47.6 $\rightarrow$ 206.171.47.7 \\ \hline
206.171.47.5  $\rightarrow$ 206.171.47.7 \\ \hline
206.171.47.2 $\rightarrow$ 206.171.47.3 \\ \hline
\end{tabular}
\end{table}

\textbf{Attacker's Profile }\\
In this section, we provide a detailed profile of the attacker.

\begin{itemize}
\item The attacker has access to some basic tools such as Nmap \cite{Nmap}, Nessus \cite{Nessus}, OpenVAS \cite{OpenVAS:scanner}, \textit{etc} and so, he can easily obtain information about the network topology and vulnerabilities information. 
\item The attacker can scan the network and discover some vulnerabilities. 
\item The attacker can find one or more hosts on the network that are having an exploitable vulnerability, as a result, he/she can perform a remote code execution attack. 
\item The attacker's goal is to execute code and gain a foothold on at-least one of the hosts, then expand access laterally through the networks until a target host is reached.
\item The attacker agent can decide the final goal automatically or decided by the security manager.
\end{itemize}

In the next section, we demonstrate the proposed framework giving the network and the attacker's profile.

\subsubsection{Phase 1}
\label{sec:Example_Data_collection}

The first phase of the framework was used to gather related information from the network, such information includes, the networked hosts, their reachability information, and the hosts' vulnerabilities and metrics. This information was collected and fed into the framework as input. Basically, 7 hosts and 27 vulnerabilities which were associated with the different version of Windows OSes were collected. The collected hosts and vulnerabilities information are shown in Table \ref{tbl:hosts_and_vuls} (some of the networked hosts have the same number and types of vulnerabilities) and one instance of the network topology is shown in Figure \ref{fig:net_topology}. 

\Figure[h!][width=0.45\textwidth]{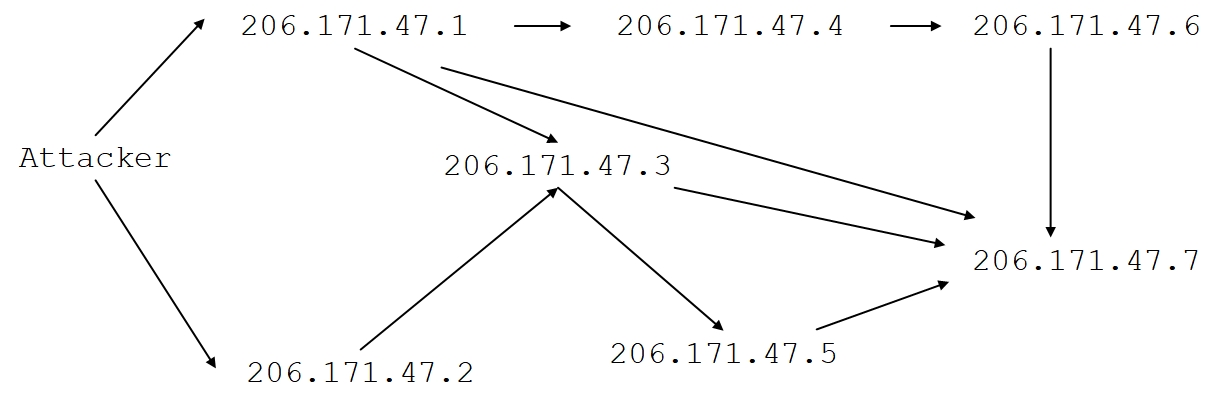}
{The Network Topology\label{fig:net_topology}}

\begin{table}[h!]
\scriptsize
\caption{The list of hosts and their vulnerabilities}
\label{tbl:hosts_and_vuls}
\begin{tabular}{l|l|c|c}
\hline
\multicolumn{4}{c}{Hosts, Vulnerabilities and Metrics} \\ \hline  \hline
\multicolumn{1}{c|}{Hosts ID} & \multicolumn{1}{c|}{Vulnerability Name} & \begin{tabular}[c]{@{}c@{}}CVSS \\ BS\end{tabular} & 
\\ \hline
\multirow{3}{*}{206.171.47.3} & \begin{tabular}[c]{@{}l@{}}DCE Services Enumeration \\ Reporting\end{tabular} & 5.0 &  \\ \cline{2-4} 
 & \begin{tabular}[c]{@{}l@{}}SMBv1 Unspecified Remote \\ Code Execution (Shadow Brokers)\end{tabular} & 10.0 &  \\ \cline{2-4} 
 & TCP timestamps & 2.6 &  \\ \hline
\multirow{2}{*}{206.171.47.1} & \begin{tabular}[c]{@{}l@{}}SMBv1 Unspecified Remote \\ Code Execution (Shadow Brokers)\end{tabular} & 10.0 &  \\ \cline{2-4} 
 & TCP timestamps & 2.6 &  \\ \hline
\multirow{6}{*}{\begin{tabular}[c]{@{}l@{}}206.171.47.4,\\ 206.171.47.5,\\ 206.171.47.7\end{tabular}} & \begin{tabular}[c]{@{}l@{}}SMBv1 Unspecified Remote \\ Code Execution (Shadow Brokers)\end{tabular} & 10.0 &  \\ \cline{2-4} 
 & \begin{tabular}[c]{@{}l@{}}DCE Services Enumeration \\ Reporting\end{tabular} & 5.0 &  \\ \cline{2-4} 
 & \begin{tabular}[c]{@{}l@{}}SSL/TLS: Certificate Signed Using \\ A Weak Signature Algorithm\end{tabular} & 4.0 &  \\ \cline{2-4} 
 & \begin{tabular}[c]{@{}l@{}}SSL/TLS: Diffie-Hellman Key\\  Exchange Insufficient DH Group \\ Strength Vulnerability\end{tabular} & 4.0 &  \\ \cline{2-4} 
 & \begin{tabular}[c]{@{}l@{}}SSL/TLS: Report Weak Cipher \\ Suites\end{tabular} & 4.3 &  \\ \cline{2-4} 
 & TCP timestamps & 2.6 &  \\ \hline
\begin{tabular}[c]{@{}l@{}}206.171.47.6,\\ 206.171.47.2\end{tabular} & TCP timestamps & 2.6 &  \\ \hline
\end{tabular}
\end{table}

\subsubsection{Phase 2}
\label{sec:Example_Gsm_construction}
Once the information from phase 1 is collected, a GSM is built and security metrics are calculated. Here, the framework was used to construct a HARM of the network, where the reachability information is captured in the upper layer and the vulnerability in the lower layer. Other information such as port number, services running can also be represented. Based on the HARM, we automatically generate possible attack paths to the target hosts (i.e., the host with IP address 206.171.47.7). The HARM with the hosts' CVSS BS score (used as risk metrics) is shown in Figure \ref{fig:harm}. \\
In this phase, we can also visualize the security model constructed and analyze the report. A screenshot of the user interface showing the HARM upper layer, the lower layer, and the security analysis report are shown in the Appendices section.\\
This phase also handles to computations and analysis of the shortest attack path and host risk.

\Figure[h!][width=0.45\textwidth]{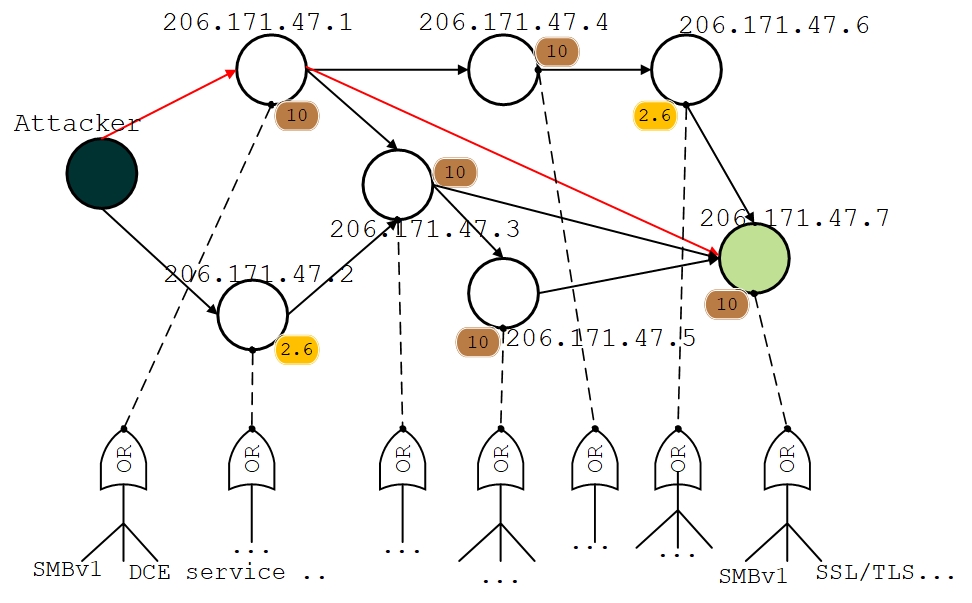}
{The HARM of the network with risk metrics \label{fig:harm}}

\subsubsection{Phase 3}
\label{sec:Example_Attack_planning}

The attack plan is the third phase of the framework. We use HARM with shortest attack path metric to plan attacker's actions, where the HARM finds all the possible set of paths to reach the target, and also calculate the shortest attack paths based on the metric calculations. Specifically, Algorithm \ref{alg:shortest_path_planning} is used for this attack planning.

Table \ref{tbl:list_paths} and Table \ref{tbl:sp_plan} shows the set of attack paths generated for the network model used, and the attack plan computed based on Algorithm \ref{alg:shortest_path_planning}, respectively.

\begin{table}[h]
\scriptsize
\caption{The list of possible attack paths generated}
\label{tbl:list_paths}
\begin{tabular}{l|l}
\hline
\multicolumn{2}{c}{The attack paths} \\ \hline  \hline 
$ap_1$ & \begin{tabular}[c]{@{}l@{}}\textit{Attacker} $\rightarrow$  \textit{206.171.47.2} $\rightarrow$ \textit{206.171.47.3} $\rightarrow$ \\ \textit{206.171.47.5} $\rightarrow$ \textit{206.171.47.7}\end{tabular} \\ \hline
$ap_2$ & \begin{tabular}[c]{@{}l@{}}\textit{Attacker} $\rightarrow$ \textit{206.171.47.1} $\rightarrow$ \textit{206.171.47.3} $\rightarrow$ \\ \textit{206.171.47.5} $\rightarrow$ \textit{206.171.47.7}\end{tabular} \\ \hline
$ap_3$ & \begin{tabular}[c]{@{}l@{}}\textit{Attacker} $\rightarrow$ \textit{206.171.47.2} $\rightarrow$ \textit{206.171.47.3} $\rightarrow$\\  \textit{206.171.47.7}\end{tabular} \\ \hline
$ap_4$ & \textit{Attacker} $\rightarrow$  \textit{206.171.47.1} $\rightarrow$  \textit{206.171.47.7} \\ \hline
$ap_5$ & \begin{tabular}[c]{@{}l@{}}\textit{Attacker} $\rightarrow$ \textit{206.171.47.1} $\rightarrow$ \textit{206.171.47.4}$\rightarrow$ \\ \textit{206.171.47.6} $\rightarrow$ \textit{206.171.47.7}\end{tabular} \\ \hline
$ap_6$ & \begin{tabular}[c]{@{}l@{}}\textit{Attacker} $\rightarrow$ \textit{206.171.47.1} $\rightarrow$ \textit{206.171.47.3} $\rightarrow$ \\ \textit{206.171.47.7}\end{tabular} \\ \hline
\end{tabular}
\end{table}

\begin{table}[h]
\caption{The attack plan generated using Algorithm \ref{alg:shortest_path_planning}}
\label{tbl:sp_plan}
\begin{tabular}{l|l}
\hline
\multicolumn{2}{l}{The attack plan} \\ \hline \hline
$ap_4$ & \begin{tabular}[c]{@{}l@{}}\textit{Attacker} $\rightarrow$ \textit{206.171.47.1}  $\rightarrow$ \textit{206.171.47.7}\end{tabular} \\ \hline
\end{tabular}
\end{table}

Since we are using only a small network with a few hosts and less network density, there is only one shortest path to the target. However, a large network may have multiple attack paths as the shortest path to the target. In this regard, the attack planner will need to select the most critical path from the list of the shortest paths by prioritizing the attack paths using attack path risk metrics. Here, the selected critical path is the attack path having the highest risk value. 


\subsubsection{Phase 4}
\label{sub:Example_Attack_execution_evaluation}

To demonstrate this phase, we utilized the Metasploit framework \cite{Metasploit} as our adversary attacking tool to attack the windows machine in the network. We choose the Metasploit because it is one of the best-known attacking tools used by cyber-criminals as well as ethical hackers to probe systematic vulnerabilities on networks. Moreover, it is an open-source framework that can easily be customized and used with most OSes. Besides, any attack tool can be developed and used with this framework.

In this phase, we fed the attack execution phase with the attack plan generated from phase 3 for execution. In particular, we customized the Metasploit to process the attacker's actions based on the output from our attack planner. Here, we utilize the Metasploit's attacks. First, the framework chooses the exploit  "exploit/windows/smb/ms17\_010\_eternalblue" and then attack the initial host -  206.171.47.7. Following that setting, the payload "payload/generic/shell\_reverse\_tcp" is used to exploit the host.

Typically, Metasploit is used as an interactive tool or repl (read, evaluate, print, loop), (i.e., meaning user type in a command, the interactive shell execute the command and print the result, and then the interactive shell waits for users to type in the next command), but for this framework, we automate the attack execution with least human interactions. Specifically, we use the resource script mechanism provided by the Metasploit, the resource script mechanism allow batching multiple Metasploit commands and execute them as one. We summarise the resource script used to attack one host as follows:\\ \\ \\

\begin{mdframed}
{\fontfamily{LinuxBiolinumT-OsF}\selectfont

\textcolor{Blue}{The target host to exploit:}\\
   \hspace*{5mm} current host = 206.171.47.7\\ 
\textcolor{Blue}{UUID generate signal:}\\
    \hspace*{5mm} succeed signal=`252b2c95-2f8a..'\\
    \hspace*{5mm} fail signal=`e59aa317-e9..'\\
\textcolor{Blue}{set the exploit module:} \\
      \hspace*{5mm} \textcolor{blue}{run\_single}(\textcolor{Bittersweet}{`use exploit/windows/smb/ms17\_010\_\\ 
      \hspace*{5mm} eternalblue'}) \\
\textcolor{Blue}{Set the target host to exploit:}\\
      \hspace*{5mm} \textcolor{blue}{run\_single}(\textcolor{Bittersweet}{`set RHOST 206.171.47.7'})\\ 
\textcolor{Blue}{Set the payload to use:} \\
\hspace*{5mm} \textcolor{blue}{run\_single}(\textcolor{Bittersweet}{`use payload generic/shell\_reverse\_tcp'})\\
\textcolor{Blue}{Starting the exploit:}\\
\textcolor{blue}{run\_single}(\textcolor{Bittersweet}{`exploit -J -z'})\\
newest\_session\_id = framework.sessions
                    .keys.max \\ \\
\textcolor{blue}{if framework}\\
   \hspace*{5mm} .sessions[newest\_session\_id] \\
   \hspace*{5mm} .target\_host == current\_rhosts \\
 \textcolor{Blue}{Show the UUID for success:}\\
   \hspace*{5mm} \textcolor{blue}{print\_line succeed\_signal}\\
   
\textcolor{blue}{else}\\
\textcolor{Blue}{Show the UUID for failure:}\\
   \hspace*{5mm}\textcolor{blue}{print\_line fail\_signal}
   
\textcolor{blue}{end}
} 
\end{mdframed}

To implement the attack in sequence, we implemented python functions to generate the resource script based on the generated attack plan (from phase 3). Since the resource script mechanism can automate only single exploitation for a single host, we use Pymetasploit3 \cite{Pymetasploit3} to automate the exploitation for a sequence of hosts on an attack path (given from the planner).  
In Figure \ref{fig:execution_vis}, we show the attack flow, and then we described the attack execution with the following steps:
\begin{enumerate}
    \item Launch Metasploit program with command \textit{msfconsole}
    \item Activate the Metasploit rpc mechanism with command \textit{load msgrpc Pass=test} within the Metasploit console interface.
    \item the Python program reads the attack plan generated from phase 3 and generates an attack path, which is just a python list data structure, then pass the python list to step 4.
    \item the Python program checks whether the python list is empty. If it is, stop. If it is not, pops out the first node in the python list, and pass that node to step 4.
    \item the Python program generates Metasploit resource script by extracting information from the node, and then via rpc make the Metasploit to start exploitation. 
    If the node is successfully exploited, a command shell is returned which can be used to control the machine, and then the next step continues based on step 3. However, if the node cannot be exploited, then, it terminates.
\end{enumerate}

\Figure[h!][width=0.23\textwidth]{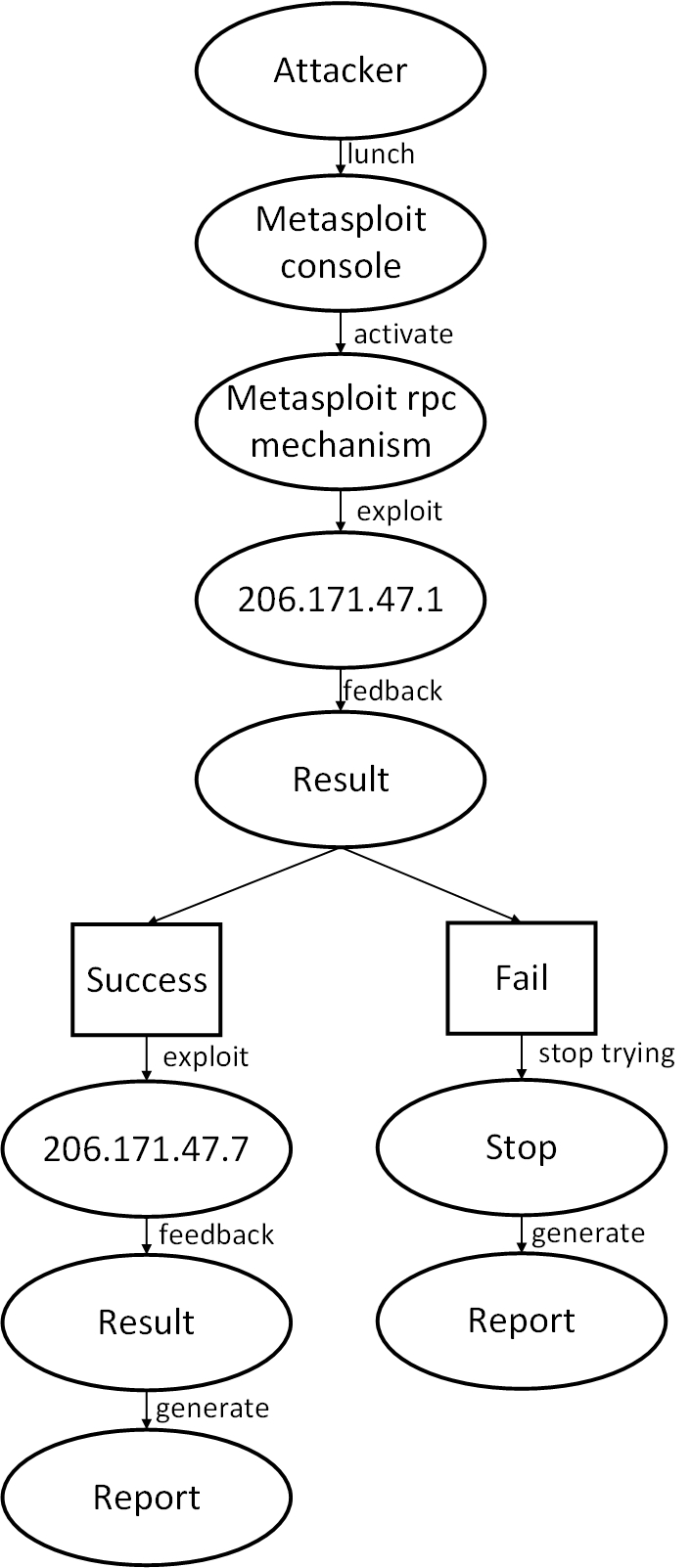}
{The attack execution and evaluation steps \label{fig:execution_vis}}

\begin{figure}[!ht]
   \centering
\resizebox{0.4\textwidth}{!}{
\begin{tikzpicture}
\begin{axis}[
    ybar,
    enlargelimits=0.08,
    scale only axis,
    ylabel={Metrics},
    symbolic x coords={AIM,ROA, High, Medium, Low, NAS},
    xtick=data,
    height=6.3cm,
    grid=both,
    nodes near coords,
    nodes near coords align={vertical},
    ]
\addplot coordinates {(AIM,162) (ROA,153.31) (High,20) (Medium,52) (Low,28) (NAS,6)};
\end{axis}
\end{tikzpicture}
}
 \caption{Case study network: security evaluation}
   \label{fig:casestudymetrics}
\end{figure}
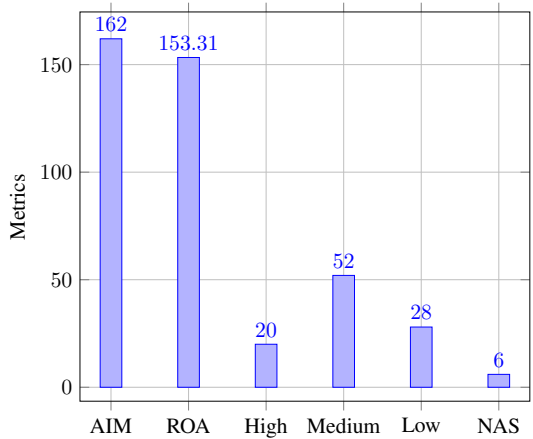

\begin{table}[!h]
\centering
\scriptsize
\caption{Vulnerabilities and attack exploit}
\label{tbl:vuls_exploits}
\begin{tabular}{l|l|l|l}
\hline
 & Number known & Used in the framework & not used \\ \hline \hline
Vulnerabilities & 25 & 18 & 7 \\ \hline
Attack exploits & 18 &  18& 0 \\ \hline
\end{tabular}
\end{table}

\begin{table}[h!]\small

{\fontfamily{LinuxBiolinumT-OsF}\selectfont

\caption{The attack report generated after executing the attack plan}
\label{tbl_Attack_report}

\begin{tabular}{l|l} 
\hline
& Attack Report \\  \hline  \hline
    & The attacker's (attacking host) ID: 192.168.1.14,   \\
    &The total number of hosts exploited: 2 \\
    & \textcolor{red}{Description}: [\{ \\
       &  \textcolor{blue}{Order of exploit on path}: Host 001, \\ 
       & \textcolor{blue}{Host ID}: 206.171.47.1, \\ 
        &\textcolor{blue}{Host running}: true, \\ 
        &\textcolor{blue}{Host exploited}: true, \\ 
       & \textcolor{blue}{CVE ID}: 2017-0143, \\ 
            & \textcolor{blue}{Exploit used}: "exploit/windows/smb/ms17\_010\_eternalblue", \\ 
           & \textcolor{blue}{The payload used}: "payload/generic/shell\_reverse\_tcp", \\ 
          &  \textcolor{blue}{Vulnerabilities associated with attack type:} CVE-2017-0143,  \\ 
          & \hspace*{3mm} CVE-2017-0144, CVE-2017-0145, CVE-2017-0146, \\
          &\hspace*{3mm} CVE-2017-0147, CVE-2017-0148, MSB-MS17-010 \\
       &\},\\
       &\\
  &   \{ \\ 
      &  \textcolor{blue}{Order of exploit on path}: Host 002, \\ 
      &  \textcolor{blue}{Host ID}: 206.171.47.7, \\ 
      &  \textcolor{blue}{Host running}: true, \\ 
       & \textcolor{blue}{Host exploited}: true, \\ 
      &  \textcolor{blue}{CVE ID}: 2017-0143, \\ 
       &    \textcolor{blue}{Exploit used}: "exploit/windows/smb/ms17\_010\_eternalblue", \\ 
         &   \textcolor{blue}{The payload used}: "payload/generic/shell\_reverse\_tcp", \\ 
         &  \textcolor{blue}{Vulnerabilities associated with attack type}: CVE-2017-0143,  \\ 
          & \hspace*{3mm} CVE-2017-0144, CVE-2017-0145, CVE-2017-0146, \\
          & \hspace*{3mm} CVE-2017-0147, CVE-2017-0148, MSB-MS17-010 \\
  &  \}] \\ 
\hline

\end{tabular}

}
\end{table}

Table \ref{tbl_Attack_report} shows the complete report on the attack executed for one attack path. Here, the report captures the host's ID, the status of the host at the time of the exploitation (i.e., whether it is running or unavailable, the status of exploit (exploited or failed), \textit{etc}.

Table \ref{tbl:vuls_exploits} shows the total number of possible attacks found and also the total number of corresponding exploit modules for the attacks. In Figure \ref{fig:casestudymetrics}, we show the security evaluation for the case study network, Where there is a total of 6 attack scenarios to reach the host 206.171.47.1, and the overall attack impact on the network is calculated as 162. This indicates this impact of the damages as a result of the attack on the overall network. As the decision-maker, it is expected that security defenses should be deployed such as the impact metric is low. The ROA metric shows the expected benefit for the attacker to compromise the network system and it is calculated at 153.31. In addition, the severity of the attacks is shown by High, Medium, and Low, where 52\% of the network hosts have a medium level of exploitable attacks (or vulnerability severity), and the others are 20\% and 28\% for high and low, respectively.

\section{Experiments and Analysis}
\label{sec:Experiments}
We perform real experiments based on a commercial cloud- Amazon Web Services (AWS) \cite{Amazon} on two network models; (i) A three tiers network, (ii) A flat network. We use Amazon's Elastic Compute Cloud (EC2) to obtain and configure 101 virtual computer nodes (hosts) (The specifications and instances used are shown in Table \ref{tbl:resources}). We deploy our framework on one of the Amazon's host which serve as the red team agent. Next, we use the remaining 100 hosts to create a network for the experiments. 
In Section \ref{secsub_exp_setting}, we describe the network configurations used. In Section \ref{secsub_exp1}, we perform two experiments, where we investigate the time it takes to successfully process and executes each phase of the automation framework on two network models and the time to complete different attack plans. Furthermore, we include security analysis results for the experiments.

\Figure[!h][width=0.4\textwidth]{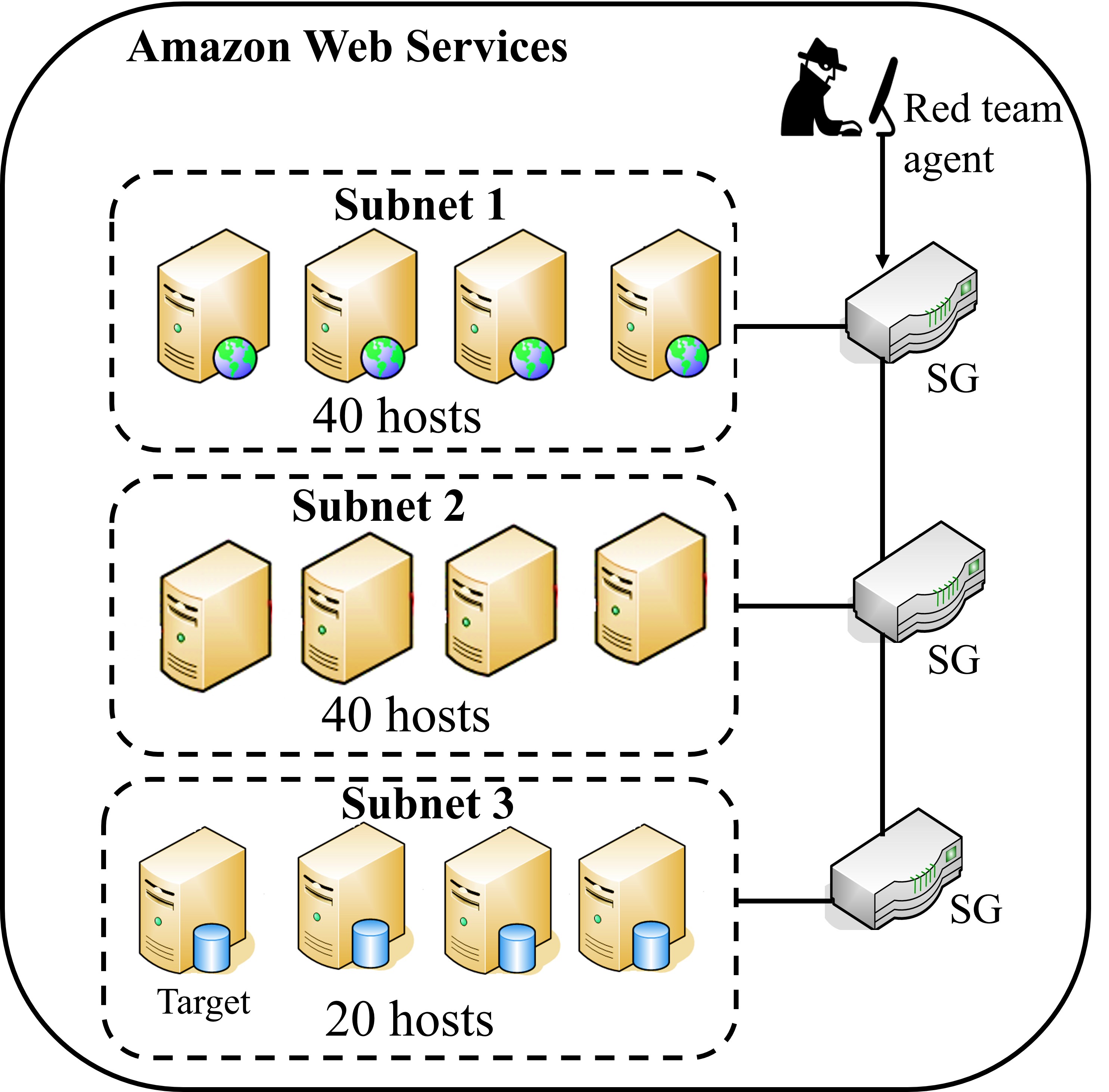}
{Three tiers network setup in the AWS \label{fig:experiment1}}

\Figure[!h][width=0.4\textwidth]{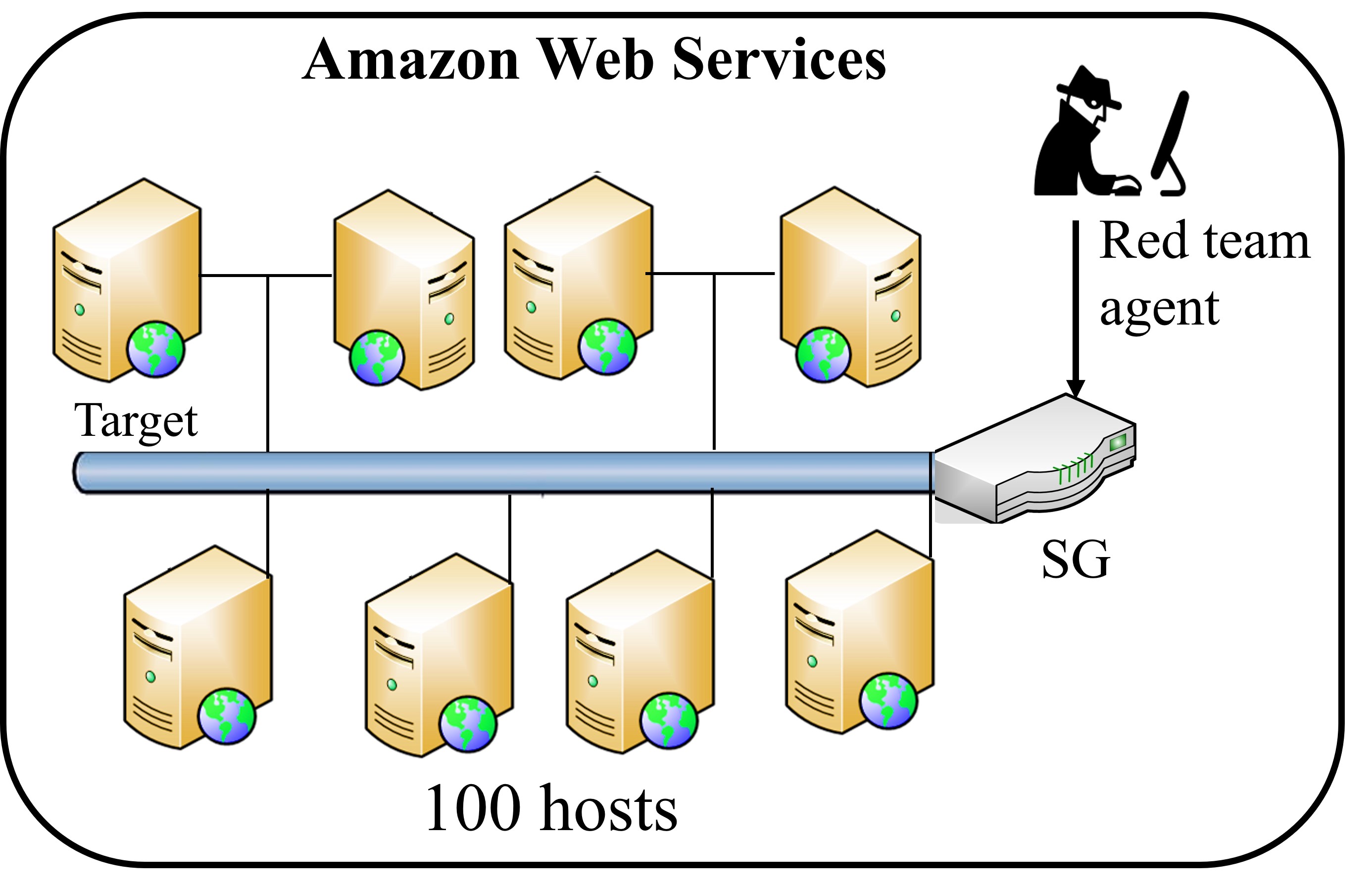}
{Flat network setup in the AWS \label{fig:experiment2}}

\subsection{Network settings}
\label{secsub_exp_setting}
The setup for the experiments is shown in Figure \ref{fig:experiment1} and Figure \ref{fig:experiment2}. In the first model, the network is divided into three subnets; subnet 1, subnet 2, and subnet 3. In the second model, the network has only one subnet where all the hosts are located. Hosts in the experiment networks have either the Windows or Linus OS.\\ 
Each network model has Security Groups (SG) which specifies the access to host IP, ports, or subnets. The specifications in the SG is provided in Table \ref{tbl:SGs}, where SG 1 is represented by subnet 1, \textit{etc}.
The SG rules allow users outside the network to reach hosts in subnet 1 (in which the attacker is also a user), and the users cannot directly connect to hosts in subnet 2 or subnet 3. Hosts in subnet 1 are allowed to connect to hosts subnet 2, and also hosts in subnet 2 are allowed connection to hosts in subnet 3. For the second model, the SG rules allow all connections from the outside to a few hosts in the network before allowing access to other hosts in the network.

\begin{table}[h]
\caption{AWS resources for the experiments}\scriptsize
\label{tbl:resources}
\begin{tabular}{l|l|l|l}
\hline
No. & \begin{tabular}[c]{@{}l@{}}Instance \\ type\end{tabular} & Images & Specifications \\ \hline \hline
\begin{tabular}[c]{@{}l@{}}1 \\ (RT agent host)\end{tabular} & t2.medium & \begin{tabular}[c]{@{}l@{}}Ubuntu \\ 18.04 LTS\end{tabular} & \begin{tabular}[c]{@{}l@{}}Mem:4GB, \\ vCPU:2\end{tabular} \\ \hline
98 & t2.micro & \begin{tabular}[c]{@{}l@{}}Amazon \\ Linux 2\end{tabular} & \begin{tabular}[c]{@{}l@{}}Mem:1GB, \\ vCPU:1\end{tabular} \\ \hline
2 & t2.micro & Windows Server 2008 & \begin{tabular}[c]{@{}l@{}}Mem:1GB, \\ vCPU:1\end{tabular} \\ \hline
\end{tabular}
\end{table}

\begin{table}[!h]
\caption{The security groups access control rules}
\scriptsize
\label{tbl:SGs}
\begin{tabular}{l|l||l|l}
\hline
\multicolumn{2}{l||}{Three tiers network (Fig. \ref{fig:experiment1})} & \multicolumn{2}{l}{Flat network (Fig. \ref{fig:experiment2})} \\ \hline \hline
Host in & Accept from & Host & Accept from \\ \hline
Subnet 1 & All &10.50.16.73 \& & All \\ \cline{1-2}
Subnet 2 & Subnet 1 &10.50.16.82 & \\ \cline{1-2} \hline
Subnet 3 & Subnet 2 &All  & 10.50.16.73, 10.50.16.82  \\ \hline
\end{tabular}
\end{table}


\textbf{Attacker's goal}\\
We assume some of the hosts have sensitive information and the attack goal is for the attacker to reach the hosts and escalate privileges, then steal information on the host. \\ We use different targets in the experiments. For the three-tier network, we performed experiments with (a) when the target is located subnet 1, (b) when the target is located in subnet 2, and (c) when the target is located subnet 3, as shown in Table \ref{tbl:Exe_phase3}. For network model 2, we use two scenarios; (a) the target is having a Linus OS, and (b) the target is having a windows OS.

\subsection{The Experiments}
\label{secsub_exp1}
Using the network configuration in Section \ref{secsub_exp_setting}, we perform experiments to measure the time to process and execute each phase of the framework. We use two network model with the same number of hosts in the Amazon AWS, and the setup for these networks are shown in Figure \ref{fig:experiment1} and Figure \ref{fig:experiment2}. 

\subsubsection{Analysis of Phase 1 }
The framework is used to collect data from the network on Amazon EC2, where OpenVAS and the SG inbound traffic are utilized to automatically collect vulnerabilities and host reachabilities, respectively. The SG rules are summarized in Table \ref{tbl:SGs} and the vulnerabilities information are provided as supplementary material with this paper. Here, both network models have the same hosts and vulnerabilities, hence we used the same results collected from the vulnerability scanner as the input in the experiments. From the experiments, a total of 2406 vulnerabilities were collected from 100 hosts using a fast and full scan, out of which 1565 have CVE ID and 841 have no CVE ID. In these experiments, only vulnerabilities with CVE ID are used.\\
Metrics values are automatically extracted from the hosts' vulnerabilities. In particular, this framework extracts the CVSS BS of vulnerabilities that have a CVE ID (the CVSS BS provides the severity of each vulnerability  \cite{CVSS} with 10 being the most severe). We automatically assign the probability of attack success for each vulnerability based on the CVSS values (i.e., the \textit{CVSS BS score/10}). Since the CVSS does not have attack cost metrics, we also compute and assign attack cost value to the vulnerabilities based on their severity scores (i.e., a vulnerability with high severity value is assigned low attack effort costs and vice versa).

In this phase, we measure the time to collect and process the vulnerabilities and reachability information via the framework, and the results are shown in Table \ref{tbl:Exe_phase1And2}.\\ 
The results show that the framework is able to completely scan for 100 hosts in about 2hrs for the fast and full option. This scanning time is fast compared to other scanning options available (e.g., the full and very deep ultimate). In addition, the scanning time can be reduced further by reducing the port range or using other scanning options (such as discovery), however, the number of vulnerabilities captured may not be comprehensive.
On the other hand, the framework was able to process and populate the reachability information and vulnerability information within a few seconds each. Here, the reachability information was processed based on the SG collected by considering the inbound rules. 
\begin{table}[!h]
\centering
\caption{Phase 1 and Phase 2: Average execution time }
\scriptsize
    \label{tbl:Exe_phase1And2}
\begin{tabular}{l|l|c|l|l|c|l}
\hline
\multicolumn{2}{l|}{Framework Phase} & \multicolumn{3}{c|}{Three- tiers network} & \multicolumn{2}{l}{Flat network} \\ \hline \hline
\multirow{3}{*}{Phase 1} & \begin{tabular}[c]{@{}l@{}}Vulnerability\\ Scanning\end{tabular} & \multicolumn{5}{c}{2 hrs 1min 50 sec} \\ \cline{2-7} 
 & \begin{tabular}[c]{@{}l@{}}Processing \\ the vulnerabilities\end{tabular} & \multicolumn{5}{c}{0.2070 sec} \\ \cline{2-7} 
 & \begin{tabular}[c]{@{}l@{}}Processing \\ security groups\end{tabular} & \multicolumn{3}{c|}{0.0250 sec} & \multicolumn{2}{c}{0.0221 sec} \\ \hline
Phase 2 & \begin{tabular}[c]{@{}l@{}}HARM \\ construction\end{tabular} & \multicolumn{3}{c|}{7.3929 sec.} & \multicolumn{2}{c}{7.1809 sec} \\ \hline
\end{tabular}
\end{table}

\subsubsection{Analysis of Phase 2 }
Based on the output from Phase 1, a HARM is constructed and the set of possible attack paths are generated for the different network models. 
Here, we focused on measuring the time to construct HARM for each of the network models. The results are shown in Table \ref{tbl:Exe_phase1And2}. \\
The results show that the HARM for the network models was built within a few seconds for both experiments. However, the flat network took less time to build compared to the three-tiers network. This because the flat network has less network density compared to the flat network.

\subsubsection{Analysis of Phase 3 }
In this section, we analyze the time to generate attack plans using the three proposed metric-based approach. These approaches are (i) Atomic metric approach (attack cost), (ii) Path-based approach (shortest path metric), and Composite metrics (Probability of attack success on paths). We measure the time to generate an attack plan for each strategy. The results are shown in Table \ref{tbl:Exe_phase3}. \\
The results is shown in Table \ref{tbl:Exe_phase3}. For each network model, we changed the location of the target host then compute an attack plan for the different strategies. The results show that the framework computes the attack plan within a minimal time as shown in Table \ref{tbl:Exe_phase3}. Generally, we observe that the composite metrics approach takes more time to generate compared to other strategies, and depending on the density of the network, the time to compute the attack plan can vary for all planning strategies.

\begin{table*}[!h]
\centering
\scriptsize
\caption{{Phase 3: Average execution time (seconds)}}
    \label{tbl:Exe_phase3}
\begin{tabular}{c|l|c|c|c|c|c}
\hline
 & \multicolumn{1}{c|}{} & \multicolumn{3}{c|}{Three- tiers network} & \multicolumn{2}{c}{Flat network} \\ \cline{3-7} 
\multirow{-2}{*}{Phase} & \multicolumn{1}{c|}{\multirow{-2}{*}{Approach}} & \begin{tabular}[c]{@{}c@{}}Target in \\ subnet 1\\ (IP: 10.50.16.77)\end{tabular} & \begin{tabular}[c]{@{}c@{}}Target in \\ subnet 2\\ (IP: 10.50.17.117)\end{tabular} & \begin{tabular}[c]{@{}c@{}}Target in \\ Subnet 3\\ (IP: 10.50.18.99)\end{tabular} & \begin{tabular}[c]{@{}c@{}}Target 1\\ (Linux)\end{tabular} & \begin{tabular}[c]{@{}c@{}}Target 2\\ (Windows OS)\end{tabular} \\ \hline \hline
 & \begin{tabular}[c]{@{}l@{}}atomic metric \\ (attack cost)\end{tabular} & {\color[HTML]{000000} 0.20284} & {\color[HTML]{000000} 0.1894} & {\color[HTML]{000000} 3.6331} & 0.0200 & 0.0500 \\ \cline{2-7} 
 & \begin{tabular}[c]{@{}l@{}}Path-based\\ (Shortest path)\end{tabular} & {\color[HTML]{000000} 0.0944} & {\color[HTML]{000000} 0.1332} & {\color[HTML]{000000} 3.8298} & 0.0490 & 0.0320 \\ \cline{2-7} 
\multirow{-3}{*}{Phase 3} & \begin{tabular}[c]{@{}l@{}}Composite \\ metric (Prob.)\end{tabular} & {\color[HTML]{000000} 3.6125} & {\color[HTML]{000000} 3.9341} & {\color[HTML]{000000} 4.6460} & 0.0820 & 0.1030 \\ \hline
\end{tabular}
\end{table*}

\subsubsection{Analysis of Phase 4 }
The fourth phase is the attack executions and evaluation. Here, based on the different attack plans generated from Phase 3, we measure and compare the time it takes to execute an attack plan on the two network models. In Table \ref{tbl:Exe_phase4}, we show the results of this experiments.\\
Similarly, the results show that the attack executions took a few seconds to complete, with the attacks on the target which is located in subnet 3 taking more time to complete compared to target in subnet 2. Likewise, the attacks on the target that is located in subnet 2 took more time to complete compared to the target in subnet 1. This is because the number of steps required to reach a target varies for each of the subnets, where the target in subnet 3 having the highest number of steps to be reached reach. Similarly, the attacks on the flat network finished within a few seconds for target 1. However, on careful analysis of the time taken to complete attacks via the various paths, we observed that paths having the Windows OS (i.e., the target 2 scenario) takes much time to complete compared to the other scenarios with only the Linus OSes. So, the diversity of OSes on attack paths increases the time to complete the attacks. Moreover, we found that the Windows OSes have more vulnerabilities than the Linus OS, as a result, this may increase the time to exploit the vulnerabilities as the attack will make many attempts before getting a vulnerability that works.

\begin{table}[!h]
\caption{Vulnerabilities and attack exploit for the Amazon's three tier and flat network}
\scriptsize
\label{tbl:Amazonvuls_exploits}
\begin{tabular}{l|l|l|l}
\hline
 & Number known & Used in the framework & not used \\ \hline  \hline
Vulnerabilities & 2406 & 1565 & 841 \\ \hline
Attack exploits & 1565 &  1565& 0 \\ \hline
\end{tabular}
\end{table}

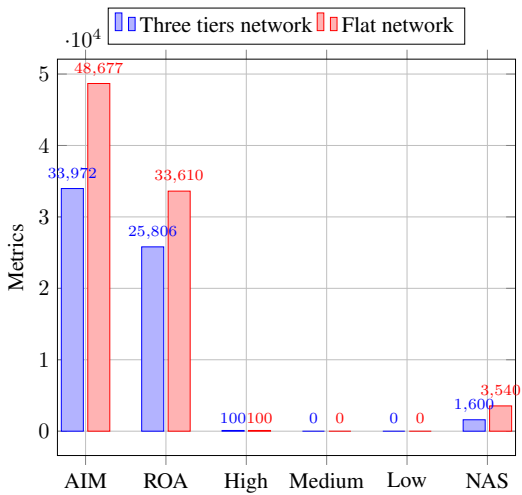
\begin{figure}[!ht]
   \centering
\resizebox{0.4\textwidth}{!}{
\begin{tikzpicture}
\begin{axis}[
    ybar,
    enlargelimits=0.07,
    scale only axis,
    legend columns=2, legend style={cells={anchor=west}, at={(0.5,1.13)}, anchor=north, draw=none, name=legend_name,draw} ,
    ylabel={Metrics},
    symbolic x coords={AIM,ROA, High, Medium, Low, NAS},
    xtick=data,
    height=6.3cm,
    grid=both,
    nodes near coords,
    nodes near coords style={font=\scriptsize},
    ]
\addplot coordinates {(AIM,33972) (ROA,25806) (High,100) (Medium,0) (Low,0) (NAS,1600)};
\addplot coordinates {(AIM,48677) (ROA,33610) (High,100) (Medium,0) (Low,0) (NAS,3540)};
\legend{Three tiers network, Flat network}
\end{axis}
\end{tikzpicture}
}
 \caption{Amazon EC2 networks: security evaluation}
   \label{fig:Amazonmetrics}
\end{figure}

In Table \ref{tbl:Amazonvuls_exploits}, we present the number of metrics vulnerabilities and exploits used. Vulnerabilities without exploits modules are not used in the framework. Hence all possible attacks executed will be successful. In Figure \ref{fig:Amazonmetrics}, we show the metrics results computed for the three tiers and flat network respectively. The framework discards vulnerabilities that it could not map to exploit modules. As a result of discarding the vulnerabilities with the no exploit modules, the \textit{Medium} and \textit{Low} is zero, and the \textit{High} is 100\% when computed against the exploitable vulnerabilities in the framework for both network models. The AIM, ROA and NAS showed high impact and attack scenario values. However, the flat network has generated a higher number of attack scenarios and high-risk values compared to the three subnet network because, the flat network has a higher network density (number of connections), and more number of attack paths.

\begin{table*}[!h]
\centering
\caption{Phase 4: Average execution time}
\scriptsize
    \label{tbl:Exe_phase4}
\begin{tabular}{c|l|l|l|l|c|c}
\hline
 & \multicolumn{1}{c|}{} & \multicolumn{3}{c|}{Three- tiers network} & \multicolumn{2}{c}{Flat network} \\ \cline{3-7} 
\multirow{-2}{*}{Phase} & \multicolumn{1}{c|}{\multirow{-2}{*}{Approach}} & \multicolumn{1}{c|}{\begin{tabular}[c]{@{}c@{}}Target in \\ subnet 1\\ (IP: 10.50.16.77)\end{tabular}} & \multicolumn{1}{c|}{\begin{tabular}[c]{@{}c@{}}Target in \\ subnet 2\\ (IP: 10.50.17.117)\end{tabular}} & \multicolumn{1}{c|}{\begin{tabular}[c]{@{}c@{}}Target in \\ Subnet 3\\ (IP: 10.50.18.99)\end{tabular}} & \begin{tabular}[c]{@{}c@{}}Target 1\\ (Linux OS)\end{tabular} & \begin{tabular}[c]{@{}c@{}}Target 2\\ (Windows OS)\end{tabular} \\ \hline \hline
 & \begin{tabular}[c]{@{}l@{}}atomic metric \\ (attack cost)\end{tabular} & {\color[HTML]{000000} 5.0545} & {\color[HTML]{000000} 8.8761} & {\color[HTML]{000000} 15.7772} & 10.0614 & 19.3675 \\ \cline{2-7} 
 & \begin{tabular}[c]{@{}l@{}}Path-based\\ (Shortest path)\end{tabular} & {\color[HTML]{000000} 4.9455} & {\color[HTML]{000000} 9.9750} & {\color[HTML]{000000} 15.5750} & 10.0147 & 20.5737 \\ \cline{2-7} 
\multirow{-3}{*}{Phase 4} & \begin{tabular}[c]{@{}l@{}}Composite \\ metric (Prob.)\end{tabular} & {\color[HTML]{000000} 5.0750} & {\color[HTML]{000000} 8.8844} & {\color[HTML]{000000} 15.7651} & 9.9501 & 20.8177 \\ \hline
\end{tabular}
\end{table*}


\section{Discussions, Limitations and Future Work}
\label{sec:discussion}

Cybersecurity modeling has been used primarily for cybersecurity analysis, evaluation, and improvement of quantitative cyber threats, attacks and defensive strategies, rather than being used for the red team (legitimate attackers) and the blue team (defenders). To automate the red team and the blue team for cybersecurity analysis, we proposed a framework to model the assessments of threats and attacks from the perspective of an attacker and the defender.
In the following sections, we discuss our work, limitations and future work.\\
\textbf{Framework implementation:}
We have used a real network to demonstrate the proposed framework. In particular, we have illustrated Phase 1 to Phase 4 of the automation framework. However, we have not implemented the defense execution and evaluation framework and workflow. It is necessary to design a defense model that provides countermeasures to cyber threats. In our future work, we will develop the defense framework and workflow.  For instance, Reed \textit{et al.} \cite{reed2014simulation} reported the defense workflow within the Cyber Security Incidence Response Teams (CSIRT) and showed how varying threats differentially affect the workflow. We can develop a similar approach for our blue team taking into account both conventional and modern defenses. Moreover, we can also consider collaborative defenders who are trying to defend the same network.\\
Furthermore, we will also need to develop a defense evaluation approach to fairly evaluate the effectiveness of the defenses \cite{Enoch:MultiObjDefs2019}. Security metrics can be developed to evaluate the effectiveness of the different defense approaches deploy, \textit{etc}.\\

\textbf{Attack Planning Strategies:} \\
The focus of this paper is to automate threats and attack executions. However, we have developed three detailed metric-based planning strategies to use with the automation framework. These planning strategies are simple and deterministic, and in the future, we need to develop more planning strategies for non-deterministic scenarios, e.g., \cite{Sarraute2011:Algorithmplanning}. 

\textbf{Multiple attackers:} 
Although we have modeled a single attacker compromising multiple hosts to reach a target, next-generation cyber-attacks can involve collaborative attackers trying to compromise the same targets, where each of the attackers may have some specialized expertise \cite{Xu:ColloborativeAttacks2009}. Besides, multiple attackers with different attack goals is another scenario not included in this paper.  So, more research is required to include collaborative and cooperative attackers, multiple attackers with multiple targets, \textit{etc}. We can also consider various attack scenarios (e.g., Distributed Denial of Service attack \cite{Prateek:DDOS2011}).\\

\textbf{Attacker Capabilities:} Our framework used the same level of behavior and capability for the attacker. However, real-world attackers can have different behaviors and capabilities which we did not take into account. As a result, this limits our proposed approach to model different kinds of attacks along with the changes in the behavior of the attacker. To extend our proposed approach, a separate component (or module) that explicitly model the changes in the attacker's behavior or the attacker's capabilities can be incorporated \cite{Ivanova:AttackerBehaviour2013,Probst:2008}. Thus, this will allow the security administrator to perform several types of security assessment activities (e.g., based on the behavior or capabilities of the attacker).\\

\textbf{Attack tool:} 
Several attack models exist that differ in their objectives and structure (e.g., the cyber kill chain \cite{Hutchins:2011CyberKillChain}). In this paper, we have used the Metasploit framework. In the future, we plan to develop our attacking tool based on typical stages of attacks and the MITRE ATT\&CK framework \cite{Mitre_framework}. We mention the MITRE ATT\&CK because it is built based on the analysis of publicly available threat reports of actively used threats by an adversary. Moreover, the ATT\&CK provides insight into the adversary's life cycle by grouping them into different levels (tactics and techniques) that the adversary tries to achieve his goals. Although the MITRE tactics and techniques do not provide any information on how the attacker might combine different techniques to accomplish his/her goals, we can develop an approach that is similar to the work of Al-Shaer \textit{et al.} \cite{Alstatistical} and Husari \textit{et al.} \cite{husari2019learning} to analyze and characterize the MITRE techniques to determine the relationships between the threat artifacts for our model.
In addition, we can also incorporate different scenarios with an attacker having partial or full knowledge of the network, \textit{etc}.

Furthermore, it is important to take into account all vulnerabilities, however, some of the vulnerabilities collected from the Amazon EC2 network have no CVE IDs but names. As a result, it was difficult to map attacks with vulnerabilities using their names in the attacking tool. For the experiments, we filtered and used only the vulnerabilities that have CVE IDs. Moreover, our framework is not able to capture unknown or zero-day attacks \cite{zhang2018network}. In the future, we plan to research the way to capture and include zero-day and unknown attacks.\\

\textbf{Attack language:} 
An attack language will provide a universal way to translate and bridge attack plan with attack execution phase regardless of the attacking tool. However, our current work does not include an attack language, but it supports a direct conversion from the planner to the Metasploit compatible format. In the future, we plan to develop a HARMer language (an attack language) to support the conversion from planner to the Metasploit compatible or any other attacking tools.\\

\textbf{Attacker learning approach and stop criteria:} 
We have demonstrated an attacker with global learning or incremental learning of the network, and specific host as the target. 
However, determining the attacker's knowledge of the network and the attacker's target or goal is difficult. As a result, there is a need to explore different scenarios of the attacker's knowledge. Besides, more stop criteria or attack goals can be incorporated, such as; 
(1) a set of IP addresses as a target,
(2)    Incremental learning until there are no new hosts,
(3)    A certain percentage of hosts are covered by the attacker,
(4) Until a specific data is found (e.g., secret information), \textit{etc}.\\


\section{Conclusion} 
\label{sec:conclusion}
In this paper, we have developed an automation framework to model cyber-threats and attack strategies in order to assess the security of real systems. We have developed a set of deterministic planning strategies using a Graphical Security Model to plan an attacker's course of actions in a networked system. Besides, we have developed detailed algorithms to automate the proposed attack planning strategies. We performed experiments on real networks and AWS networks to demonstrate the usage of the proposed automation framework. Hence, this paper developed a new automation framework that supports the modeling of the red team operations, as well as provides a security assessment tool for cyber-defenses.

\section{Appendix}
This section provides appendices. 
Figure \ref{fig:harm_upper_viz} shows the visualization for the upper layer HARM tool (of the network model). Figure \ref{fig:harm_lower_viz} shows the the visualization for lower layer HARM using ATs, and Figure \ref{fig:report_summary_viz} shows the report generated based on HARM.

\Figure[ht!][width=0.45\textwidth]{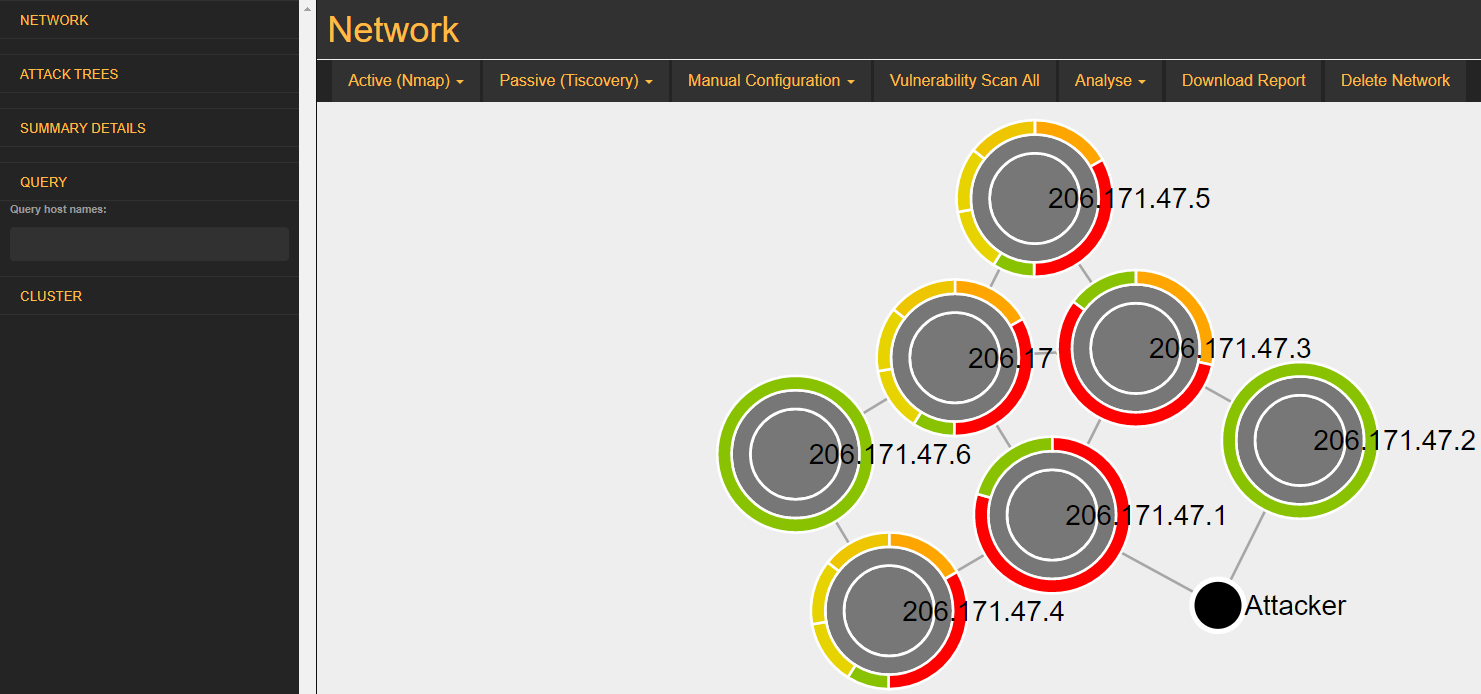}
{A screenshot of the HARM- Upper layer\label{fig:harm_upper_viz}}

\Figure[ht!][width=0.45\textwidth]{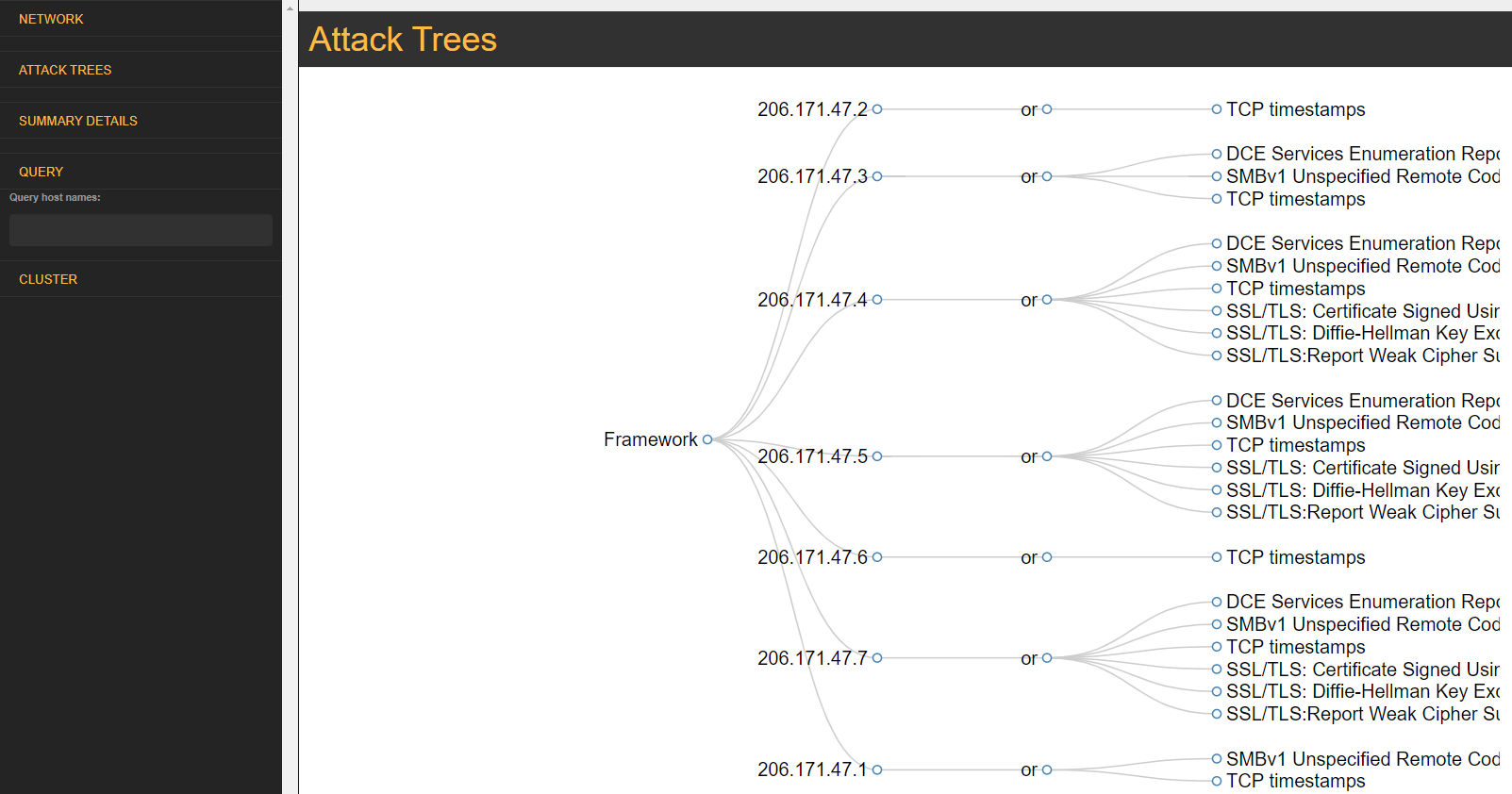}
{A screenshot of the HARM- Lower layer\label{fig:harm_lower_viz}}

\Figure[ht!][width=0.45\textwidth]{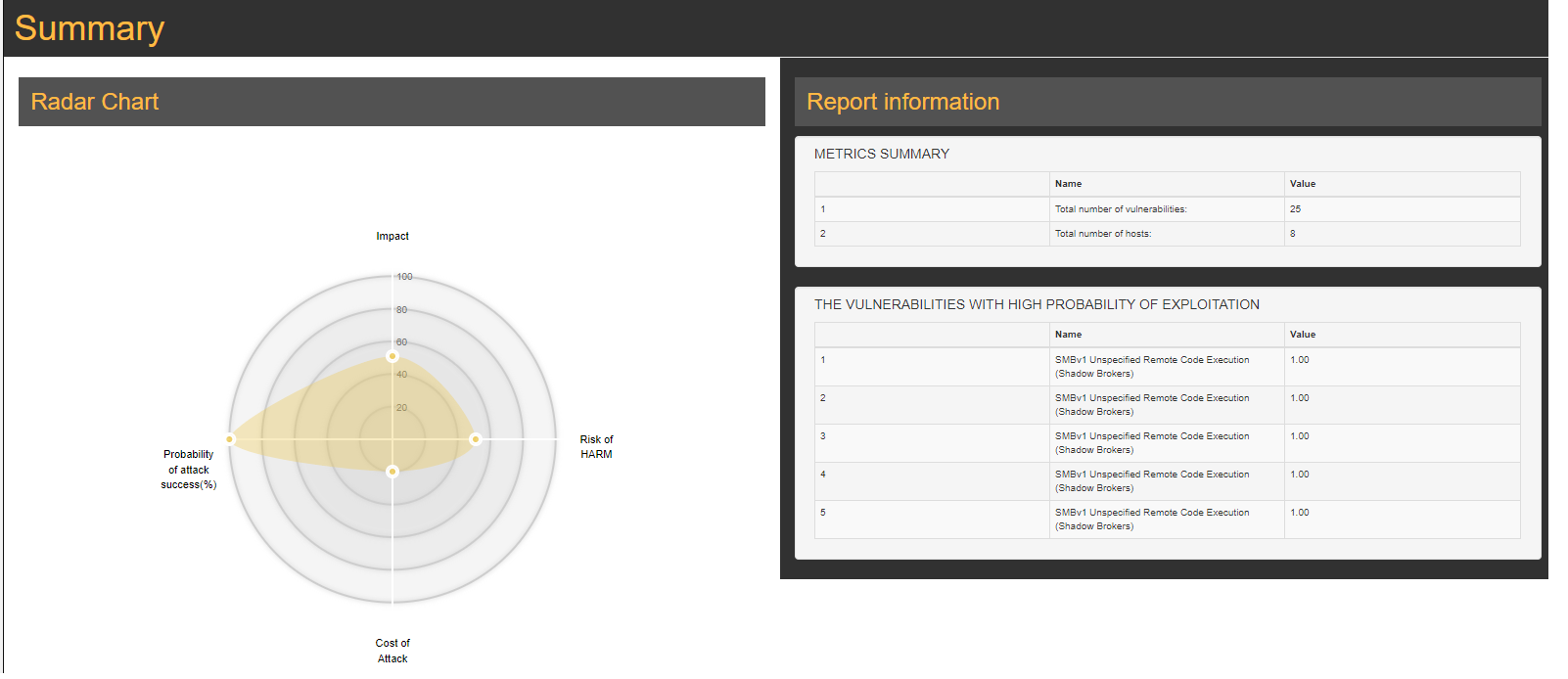}
{A screenshot of the HARM- Report\label{fig:report_summary_viz}}

\addcontentsline{toc}{section}{References}
\bibliographystyle{abbrv}
\bibliography{mybibfile}


\begin{IEEEbiography}[{\includegraphics[width=1in,height=1.25in,clip,keepaspectratio]{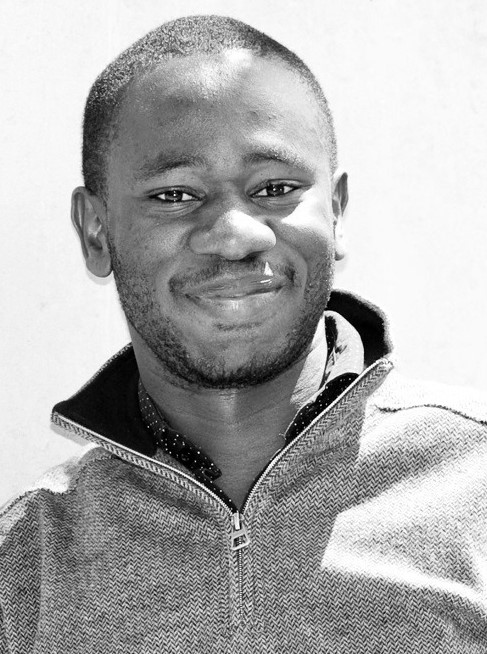}}]{Simon Yusuf Enoch (Enochson)} (Enochson) received a Ph.D. degree in Computer Science from the University of Canterbury (UC), New Zealand in 2018. He is a Postdoctoral Research Fellow with the School of Information Technology and Electrical Engineering, The University of Queensland (UQ), Australia, where he is being mentored by Assoc. Prof. Dong Seong Kim. Prior to UQ, Dr. Enochson was a Research Assistant with the Cybersecurity Research Lab. at UC, New Zealand from 2017 to 2019. He worked as an Assistant Lecturer in Computer Science with the Federal University Kashere, Gombe, Nigeria. Dr. Enochson has published papers in reputable Conferences and top-tier Journals.  His research interests include cyber-attacks \& defense automation, security modeling, and analysis of computers and networks including moving target defense.
\end{IEEEbiography}

\begin{IEEEbiography}[{\includegraphics[width=1in,height=1.25in,clip,keepaspectratio]{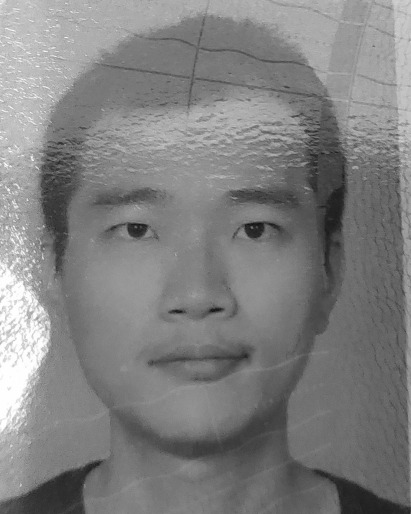}}]{Zhibin Huang} received the Bachelor of Science degree from 
the Sun Yat-sen University in 2013. He has two years of working experience as a Java Developer before he started his Master degree. He is currently an MSc student under the supervision of Dr Dong Seong Kim in the School of Information Technology and Electrical Engineering, The University of Queensland, Australia. His research interests include security modeling, penetration testing and cyber-attack automation.
\end{IEEEbiography}

\begin{IEEEbiography}[{\includegraphics[width=1in,height=1.25in,clip,keepaspectratio]{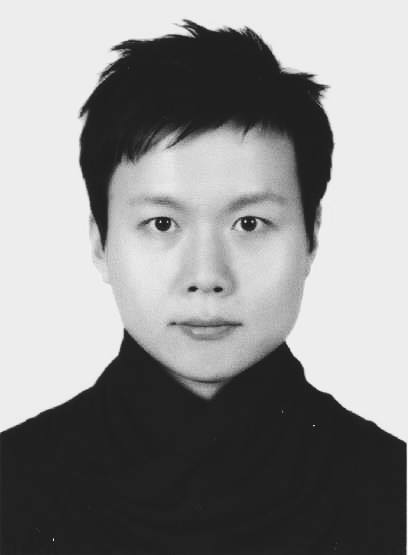}}]{Chun Yong Moon} received the B.S. in physics from Yonsei University, Korea, in 1997 and M.S degrees in computer science and engineering from the University of New South Wales, Australia in 2012, and civil engineering from The University of Queensland, Australia in 2019.
Currently, he is a Research Assistant and a PhD student under the supervision of Dr Dongseong Kim in the School of Information Technology and Electrical Engineering, The University of Queensland, Australia. His research interests include security modeling and the analysis of computers and networks.
\end{IEEEbiography}

\begin{IEEEbiography}[{\includegraphics[width=1in,height=1.25in,clip,keepaspectratio]{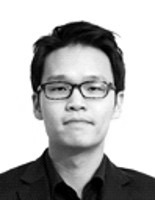}}]{Donghwan Lee}  received B.E degree in industrial engineering and M.S degree in computer science and engineering from Korea University, Seoul, Korea, in 2006 and 2008, respectively. He is currently a senior researcher at 2nd R\&D Institute, Agency for Defense Development, Seoul, Rep. of Korea. His research interests include wireless communication, parallel and distributed computing, wireless security and virtualization technologies for cybersecurity.
\end{IEEEbiography} 

\begin{IEEEbiography}[{\includegraphics[width=1in,height=1.25in,clip,keepaspectratio]{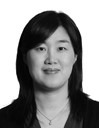}}]
{Myung Kil Ahn} received B.S. degree in information and communication engineering from Chungnam National University, Daejeon, Rep. of Korea, in 1997, and M.S. degree in computer engineering from the Sogang University, Seoul, Rep. of Korea in 2003. She is currently pursuing her Ph.D. degree in electrical and electronics engineering at Chung-Ang University. She is currently a principal researcher at 2nd R\&D institute, Agency for Defense Development, Seoul, Rep. of Korea. Her research interests include computer security and cyberwarfare modeling \& simulation.
\end{IEEEbiography}

\begin{IEEEbiography}[{\includegraphics[width=1in,height=1.25in,clip,keepaspectratio]{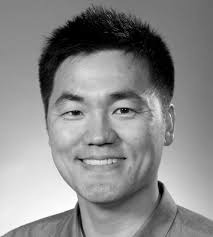}}]{Dong Seong Kim} is an Associate Professor at the University of Queensland (UQ), Brisbane, Australia. Prior to UQ, he led the Cybersecurity Lab. at the University of Canterbury (UC), Christchurch, New Zealand from August 2011 to Jan 2019. He was a Senior Lecturer in Cyber Security in the Department of Computer Science and Software Engineering at the UC. He received a Ph.D. degree in Computer Engineering from the Korea Aerospace University in February 2008. He was a visiting scholar at the University of Maryland, College Park, Maryland in the US during the year 2007 in Prof. Virgil D. Gligor Research Group. From June 2008 to July 2011, he was a postdoc at Duke University, Durham, North Carolina in the US in Prof. Kishor S. Trivedi. His research interests are in security and dependability for systems and networks; in particular, Intrusion Detection using Data Mining Techniques, Security and Survivability for Wireless Ad Hoc and Sensor Networks and Internet of Things, Availability and Security modeling and analysis of Cloud computing, and Reliability and Resilience modeling and analysis of Smart Grid.
\end{IEEEbiography}


\EOD
\end{document}